\documentstyle[aps,eqsecnum,epsf]{revtex}

\ifx\epsffile\undefined
\def\insertfig#1{}
\else
\def\insertfig#1{{\baselineskip=4pt
\centerline{\epsfxsize=\hsize\epsffile{#1}}}}\fi

\begin{document}

\def\Tr{{\rm Tr}}
\def\N{N_c}
\def\ls{\Lambda \Sigma^0}
\def\ssqr#1#2{{\vbox{\hrule height #2pt
      \hbox{\vrule width #2pt height#1pt \kern#1pt\vrule width #2pt}
      \hrule height #2pt}\kern- #2pt}}
\def\sqr{\mathchoice\ssqr8{.4}\ssqr8{.4}\ssqr{5}{.3}\ssqr{4}{.3}}

\def\bsqr{\ssqr{10}{.1}}

\twocolumn[       
{\tighten
\preprint{\vbox{
\hbox{UCSD/PTH 96--08}
\hbox{hep-ph/9603449}
}}
\title{Heavy Baryon Masses in the 1/$\bf m_Q$ and 1/$\bf N_c$ Expansions}
\author{Elizabeth Jenkins\footnotemark\footnotemark}
\address{Department of Physics, University of California at
San Diego, La Jolla, CA 92093}
\bigskip
\date{March 1996}
\maketitle
\widetext
\vskip-1.5in
\rightline{\vbox{
\hbox{UCSD/PTH 96--08}
\hbox{hep-ph/9603449}
}}
\vskip1.5in
\begin{abstract}
The masses of baryons containing a single heavy quark are studied in a combined
expansion in $1/m_Q$, $1/N_c$ and $SU(3)$ flavor symmetry breaking.
Heavy quark baryon mass splittings are related to mass splittings of the octet
and decuplet baryons.  The $\Sigma_c^*$, $\Xi_c^\prime$ and $\Omega_c^*$ are
predicted to the level of a few MeV.  A number of bottom baryon mass splittings
are predicted very accurately.
\end{abstract}

}
] 

\narrowtext

\footnotetext{${}^\dagger$Alfred P. Sloan Fellow.}
\footnotetext{${}^*$National Science Foundation Young Investigator.}

\section{Introduction}

The mass spectrum of baryons containing a single heavy quark is tightly
constrained by the presence of approximate symmetries.  In the heavy quark
limit, hadrons containing a single heavy quark respect a heavy quark
spin-flavor symmetry \cite{iw}.  For finite $m_Q$, this symmetry is broken by
effects of order $1/m_Q$.  In the large $N_c$ limit, baryons with an
approximate flavor symmetry possess a larger spin-flavor symmetry \cite{dm}.
For finite $N_c$, this symmetry is broken by effects of order
$1/N_c$\cite{dm,j,djm,lm,cgo,djmtwo,jl}.  The combined limit 
$m_b \rightarrow \infty$, $m_c \rightarrow \infty$, $N_c \rightarrow \infty$
for fixed $(m_c /m_b)$ and $(N_c \Lambda_{\rm QCD}/m_b)$ results in
a light quark and heavy quark spin-flavor symmetry $SU(6)_\ell \times SU(4)_h$
for baryons containing a single heavy quark.  For finite $m_Q$ and $N_c$, this
symmetry is violated by effects of order $1/N_c$ and $(1/N_c m_Q)$, and by
$SU(3)$ flavor breaking.  It is the purpose of this work to explore the
implications of $SU(3)$ flavor breaking and the combined $1/m_Q$ and
$1/N_c$ expansions for the masses of heavy quark baryons.   

Section~II begins with a brief review of the $1/m_Q$ expansion of heavy hadron
masses in heavy quark effective theory.  Section~III presents the operator
analysis of the $1/N_c$ expansion.  Heavy baryon masses are analyzed in the
combined $1/m_Q$ and $1/N_c$ expansion for isospin flavor symmetry in
Section~IV.  The generalization to $SU(3)$ flavor symmetry is performed in
Section~V.  The most precise predictions follow from the $SU(3)$ mass relations
of Sect.~V.  Readers interested only in these results may go directly to
subsection~VA  where they are presented.  
Isospin-violating mass splittings are considered in Section~VI.
Conclusions are presented in Section~VII.

\section{Heavy Quark Baryon Masses in the $1/m_Q$ Expansion}

The masses of hadrons containing a single heavy quark have been studied
in the $1/m_Q$ expansion of heavy quark effective theory.  A brief review
of this expansion is presented in this section.

The HQET Lagrangian,
\begin{eqnarray}
{\cal L}_{\rm HQET} &&= \overline Q_v \left( i v \cdot D \right) Q_v 
+ \overline Q_v {{(iD)^2} \over {2 m_Q}} Q_v \nonumber\\ 
&&- Z_Q \ \overline Q_v {{g G_{\mu \nu} \sigma^{\mu \nu}} \over {4 m_Q} } Q_v
+ {\cal O} \left( {1 \over m_Q^2 } \right),
\end{eqnarray}
describes the interactions of a heavy quark $Q$ with fixed velocity $v$
inside a hadron containing a single heavy quark.  The heavy quark mass $m_Q$ is
removed from the QCD Lagrangian by the heavy quark field redefinition\cite{georgi}.  The
residual mass term has been chosen so that there is no mass term in the HQET
Lagrangian\cite{fln}.  

The mass of a hadron containing a single heavy quark has an expansion in $1/m_Q$, 
\begin{equation}
M\left(H_Q\right) = m_Q + \bar \Lambda -{{\lambda_1} \over {2m_Q}} 
- d_H {{\lambda_2} \over {2m_Q} }
+ {\cal O}\left({1 \over m_Q^2} \right),
\end{equation}
where the order unity contribution $\bar \Lambda$ is the mass of the light degrees
of freedom in the hadron, and the two $1/m_Q$ contributions are determined by
the matrix elements 
\begin{equation}
\lambda_1 = \left<{H_Q(v)}\left|
\overline Q_v \left( i D \right)^2 Q_v 
\right|{H_Q(v)}\right>
\end{equation}
and
\begin{equation}
d_H \lambda_2 = \frac 1 2 Z_Q\ \left< H_Q(v) \left| 
\overline Q_v g G_{\mu \nu} \sigma^{\mu \nu} Q_v 
\right| H_Q(v) \right>.
\end{equation}
In the above equation, $d_H$ is the Clebsch factor
$d_H = -4 \left( J_\ell \cdot J_Q \right) $, and $Z_Q$ is a renormalization
factor with $Z_Q(\mu = m_Q) = 1$.  Renormalization group scaling between the
scales $m_b$ and $m_c$ yields
$Z_b/Z_c = \left[ \alpha_s(m_b)/ \alpha_s(m_c) \right]^{9/25}$.

Neglecting $SU(3)$ flavor breaking,
the masses of the lowest-lying pseudoscalar and vector mesons $D$ and $D^*$ for
$Q=c$ and $B$ and $B^*$ for $Q=b$ are parametrized by
\begin{eqnarray}
P_Q &&= m_Q + \bar \Lambda^{\rm meson}- { \lambda_1^{\rm meson} \over {2m_Q}}
- {{3\lambda_2^{\rm meson}} \over {2m_Q}}+ ..., \nonumber\\
P_Q^* &&= m_Q + \bar \Lambda^{\rm meson}- { \lambda_1^{\rm meson} \over {2m_Q}}
+ {{\lambda_2^{\rm meson}} \over {2m_Q}}+ ..., 
\end{eqnarray}
whereas the masses of the lowest-lying spin-$\frac 1 2$ antitriplet ($\bar 3$)
and spin-$\frac 1 2$ and spin-$\frac 3 2$ sextets ($6$) are parametrized by 
\begin{eqnarray}
T_Q &&= m_Q + \bar \Lambda^{\rm baryon}_T
- { \lambda_{1T}^{\rm baryon} \over {2m_Q}}
+ ..., \nonumber\\
S_Q &&= m_Q + \bar \Lambda^{\rm baryon}_S
- { \lambda_{1S}^{\rm baryon} \over {2m_Q}}
- {{4\lambda_{2S}^{\rm baryon}} \over {2m_Q}}+ ..., \\
S_Q^* &&= m_Q + \bar \Lambda^{\rm baryon}_S
- { \lambda_{1S}^{\rm baryon} \over {2m_Q}}
+ {{2\lambda_{2S}^{\rm baryon}} \over {2m_Q}}+ ..., \nonumber
\end{eqnarray}
to order $1/m_Q^2$ in the $1/m_Q$ expansion.  Note that different parameters
$\bar \Lambda$, $\lambda_{1}$ and $\lambda_2$
appear for mesons and baryons. 
When $SU(3)$ flavor breaking is not neglected, different
parameters also appear for different members of the meson and baryon $SU(3)$ 
multiplets.

\section{Operator Analysis}

The $1/N_c$ operator analysis for baryons
containing light quarks
was derived in detail in Ref.~\cite{dm,j,djm,lm,cgo,djmtwo}.  The operator
analysis is generalized to baryons containing heavy quarks in this section.

The approximate flavor symmetry of QCD with three light
quarks $u,d,s$ and heavy quarks $c$ and $b$ is $SU(3) \times U(1)_c \times
U(1)_b$.  The corresponding
spin-flavor symmetry for large-$N_c$ baryons is  
$SU(6) \times SU(2)_c \times SU(2)_b$.
There is a separate spin symmetry for each flavor of heavy
quark in the baryon in the large-$N_c$ limit.  
This heavy quark spin symmetry is operative for baryons
containing multiple heavy quarks as well as baryons containing a single heavy
quark.  It is important to emphasize that large-$N_c$ heavy quark spin
symmetry is present for baryons even if
the quark flavor in question is not heavy enough to have a valid $1/m_Q$
expansion.  For this reason, it is possible to study the consequences of 
large-$N_c$ heavy quark spin symmetry for baryons in the $1/N_c$ expansion
without reference to the $1/m_Q$ expansion.  We
begin with this $1/N_c$ operator
analysis and then generalize to the combined $1/m_Q$ and
$1/N_c$ expansion.   
 
The operator basis of the $1/N_c$ expansion for baryons containing heavy quarks
is constructed out of the zero-body identity operator $\openone$ and the
generators of the spin-flavor algebra $SU(6) \times SU(2)_c \times SU(2)_b$.  The $SU(6)$
generators are defined by  
\begin{eqnarray}
&&J_\ell^i = q^\dagger {\sigma^i \over 2} q, \nonumber\\
&&T^a = q^\dagger {\lambda^a \over 2} q, \nonumber\\
&&G^{ia} = q^\dagger {{\sigma^i \lambda^a} \over 4} q, \nonumber
\end{eqnarray}
where $J_\ell^i = J_u^i + J_d^i + J_s^i$, while the charm and bottom spin
generators are defined by
\begin{eqnarray}
&&J_c^i = c^\dagger {\sigma^i \over 2} c, \nonumber\\ 
&&J_b^i = b^\dagger {\sigma^i \over 2} b. 
\end{eqnarray}
The light quark spin-flavor algebra is given in Ref.~\cite{djmtwo}.  The
heavy quark spin symmetry commutation relations are simply 
\begin{equation}
\left[ J_Q^i, J_Q^j \right] = i \epsilon^{ijk} J_Q^k,
\end{equation}
where $Q=c$ or $b$.

Operators in the $1/N_c$ expansion are polynomials in the $1$-body operators.
The set of polynomial operators is overcomplete.  An operator basis which is
complete and linearly independent is constructed by eliminating redundant
operators using operator identities.  The operator identities are derived in
Ref.~\cite{djmtwo}.  The structure of the operator identities is discussed in
detail in Ref.~\cite{djmtwo}, and will not be repeated here.  For the case of
baryons containing $N_\ell$ light quarks and $N_h$ heavy quarks such that
$N_c = N_\ell + N_h$, the $SU(2N_F)$ identities of Ref.~\cite{djmtwo}
apply for $J^i$ replaced by $J_\ell^i$ and $N_c$ replaced by $N_\ell$ in
Eq.~(4.2) and Tables~VII and~VIII.  Similarly, the $SU(2)_Q$ spin symmetry
identities are the identities for one quark flavor with $N_c$ replaced by
$N_Q$, namely
\begin{eqnarray}
Q^\dagger Q &&= N_Q \openone, \nonumber\\
\{J_Q^i, J_Q^i\} &&= \frac 1 2 N_Q \left( {N_Q} + 2 \right) \openone.
\end{eqnarray}  
The presence of $N_\ell$ and $N_Q$ in the operator identities rather than
$N_c$ follows from the group theory of the baryon representation for a 
fixed number of light quarks and a fixed number of 
heavy quarks of each flavor.  
Fig.~1 gives the $SU(6) \times SU(2)_Q \times SU(2)_{Q^\prime}$ 
spin-flavor representation for baryons containing $N_\ell$ light quarks and 
$N_h=N_Q + N_{Q^\prime}$ heavy quarks, with $N_\ell + N_h = N_c$, where
$Q$ and $Q^\prime$ represent two different flavors of heavy quark $c$ and $b$.
It is clear that the operator identities apply for the quark numbers
$N_\ell$ and $N_Q$.
For example,
the total number of quarks in the baryon is described
by the $1$-body $\rightarrow$ $0$-body identities  
\begin{eqnarray}
q^\dagger q + Q^\dagger Q + Q^{\prime \dagger} Q^{\prime} 
&&= \left( N_\ell + N_Q + N_{Q^\prime} \right) \openone \nonumber\\
&&= \left( N_\ell + N_h \right)\openone = N_c \openone . 
\end{eqnarray}       

The $1/N_c$ expansion of a $l$-body QCD operator acting on baryons
containing $N_\ell$ light quarks and $N_h$ heavy quarks has the form 
\begin{equation}\label{nexp}
{\cal O}_{\rm QCD}^{(l)} = N_c^l \sum_{m=0}^{N_\ell} \sum_{n=0}^{N_h}
c^{(m+n)} {1 \over N_c^{m+n}} {\cal O}^{(m)}_\ell {\cal O}^{(n)}_h,
\end{equation}
where ${\cal O}_\ell^{(m)}$ denotes an $m$-body operator in the light
quark spin-flavor generators $J_\ell^i$, $T^a$, and $G^{ia}$, and ${\cal
O}_h^{(n)}$ denotes a $n$-body operator in the heavy quark 
generators\footnote{For baryons containing $N_Q$ heavy quarks of type $Q$ and
$N_{Q^\prime}$ heavy quarks of type $Q^\prime$, ${\cal O}_h^{(n)}$ is an
$n$-body operator which is at most $N_Q$-body in $J_Q^i$ and at most
$N_{Q^\prime}$-body in $J_{Q^\prime}^i$,
\begin{equation}
\sum_{p+q=0}^{N_h}{1 \over N_c^{p+q}}{\cal O}^{(p+q)}_h \rightarrow 
\sum_{p=0}^{N_Q} \sum_{q=0}^{N_{Q^\prime}} {1 \over N_c^{p+q}}
{\cal O}_{Q}^{(p)}{\cal O}_{Q^\prime}^{(q)} .
\end{equation}
}.
The $1/N_c$ operator expansion only goes up to $N_c$-body operators
since $N_c = N_\ell + N_h$.
Each of the arbitrary coefficients of the $1/N_c$ expansion
\begin{equation}
c^{(m+n)}\left( {1 \over N_c} \right)
\end{equation}  
is order one at leading order in the $1/N_c$ expansion.
The coefficients are only a function of $1/N_c$ since
$N_\ell$, $N_Q$ and $N_{Q^\prime}$ are to be regarded as fixed numbers,
not operators, in the present expansion.  The coefficients depend implicitly on
the fixed ratios $N_\ell/N_c$, $N_Q/N_c$ and $N_{Q^\prime}/N_c$.  

It is an important physical point that the operator expansion for
baryons containing light and heavy quarks is suppressed by factors of
$1/N_c$, rather than $1/N_\ell$ and $1/N_h$.  This $1/N_c$ suppression is
required for consistency of the operator expansion.  One consistency
requirement is that
the $1/N_c$ power counting of the operator expansion~(\ref{nexp}) be preserved
under commutation of the $(m+n)$-body operators 
${\cal O}^{(m+n)}\equiv {\cal O}_\ell^{(m)} {\cal O}_h^{(n)}$.  This feature is
present for a suppression factor of $1/N_c^{m+n}$, but not for $1/(N_\ell^m
N_h^n)$.    
A second consistency requirement is that the operator expansion be
invariant under a change of operator basis for the expansion.  For example, it
is possible to rewrite the operator basis using $J = J_\ell + J_h$, the total
baryon spin, rather than $J_\ell$.  This recasting of the operator expansion
is obviously possible if $J$, $J_\ell$ and $J_h$ are all suppressed by $1/N_c$,
but it is not consistent if $J_\ell$ is suppressed by $1/N_\ell$ and $J_h$ is
suppressed by $1/N_h$.  The conclusion that all the suppression factors are
$1/N_c$ also follows from an analysis of the large-$N_c$ Feynman diagrams.

Although factors of $1/N_\ell$ and $1/N_h$ do not appear as explicit
suppression factors in the $1/N_c$ operator expansion, $N_\ell$ and $N_h$ do
appear implicitly through the matrix elements of ${\cal O}_\ell$ and ${\cal
O}_h$.  Matrix elements of the operator ${\cal O}_\ell^{(m)}$ are at
most order $(N_\ell)^m$, whereas matrix elements of ${\cal O}_h^{(n)}$ are at
most $(N_h)^n$, so that 
\begin{equation}
{1 \over N_c^{m+n}}\langle{\cal O}_\ell^{(m)} {\cal O}_h^{(n)}\rangle 
\ \raise.3ex\hbox{$<$\kern-.75em\lower1ex\hbox{$\sim$}}\ 
\left({N_\ell \over N_c}\right)^m \left({N_h \over N_c}\right)^n .
\end{equation}
Both $N_\ell/N_c$ and $N_h/N_c$ are suppression factors so long as neither
$N_\ell$ nor $N_h$ is equal to $N_c$.  Thus, for
baryons containing both light and heavy quarks,
higher-body operators in the
$1/N_c$ expansion {\it are} suppressed relative to lower-body operators
if the operator basis is written in terms of light-quark and heavy-quark
$1$-body operators.  Alternative operator bases
are less ideal since they will not have this feature.  For example, the highest
spin baryon states have $J \sim N_c$, so that the use of the $1$-body operator
$J$ instead of $J_\ell$ will not lead to an operator basis with this feature.  For the physical situation of principle interest in this work, namely baryons
containing a single heavy quark, the 
$N_\ell/N_c$ and $N_h/N_c$ suppression factors will be $2/3$ and $1/3$,
respectively.

The large-$N_c$ heavy quark spin symmetry is independent of whether
the quark flavor in question is heavy enough to have a valid $1/m_Q$ expansion.
If the quark flavor is sufficiently heavy that it has a valid $1/m_Q$
expansion, then heavy quark spin symmetry is also 
a consequence of the $m_Q \rightarrow \infty$ limit, 
independent of whether the large-$N_c$ limit is taken.  It thus follows that
heavy quark spin symmetry violation is suppressed by $(1/N_c m_Q)$ for very
heavy quarks.

For the special case of baryons containing a single heavy quark, there
is a heavy quark flavor symmetry in the limit that each of the heavy quark
flavors is regarded as infinitely heavy.  More specifically, for baryons
with $N_h=1$ in the limit $m_b \rightarrow \infty$ and $m_c \rightarrow \infty$
with $m_c/m_b$ held fixed, heavy quark spin symmetry for each of the heavy
flavors is promoted to a heavy quark spin-flavor symmetry $SU(4)_h$\cite{iw}.
In the presence of heavy quark flavor symmetry (regardless of the origin of the
heavy quark flavor symmetry\footnote{ For
example, if there were two moderately heavy quarks of similar mass in QCD,
there would be a flavor symmetry amongst these quarks and a spin-flavor
symmetry for baryons containing these quarks in the large $N_c$ limit.  There
would not be a spin-flavor symmetry for mesons in this situation.}), heavy
quark spin-flavor symmetry for baryons is a consequence of the large $N_c$
limit.  Thus, when
the large-$N_c$ limit, $N_c \rightarrow \infty$ with $N_c \Lambda_{\rm
QCD}/ m_Q$ held fixed, is added to the heavy quark limit, there is a
$SU(6)_\ell \times SU(4)_h$ spin-flavor symmetry for baryons with one heavy
quark\cite{j}.  
The heavy quark spin-flavor algebra is generated by the $1$-body
operators 
\begin{eqnarray}
&&J_h^i = Q^\dagger {\sigma^i \over 2} Q, \nonumber\\ 
&&I_h^a = Q^\dagger {\tau^a \over 2} Q, \\
&&G_h^{ia} = Q^\dagger {{\sigma^i \tau^a} \over 4} Q ,\nonumber 
\end{eqnarray}
where $a=1,2,3$ is the heavy flavor quark index.  The charm and bottom quark
spin and number operators are related to the heavy quark operators and $N_h$
through 
\begin{eqnarray}
&&J_h^i = J_c^i + J_b^i, \nonumber\\ 
&&I_h^3 = \frac 1 2 \left( N_{\rm charm} - N_b \right),\\
&&G_h^{i3} = \frac 1 2 \left( J_c^i - J_b^i \right), \nonumber 
\end{eqnarray}
and $N_h = N_{\rm charm} + N_b$.  The heavy quark flavor symmetry only
applies to baryons with $N_h=1$, so operators in the $1/N_c$ expansion for
these baryons will contain no more than one heavy quark $1$-body operator
$J_h^i$, $I_h^a$ and $G_h^{ia}$.

Finally, the expansion~(\ref{nexp}) can easily
be generalized to encompass baryons with differing number of light quarks and
heavy quarks $Q$ and $Q^\prime$ by including all operators upto $N_c$-body in
the light quark $1$-body operators and the heavy quark $1$-body operators,
which now consist of $N_Q$, $J_Q^i$,
$N_{Q^\prime}$ and $J_{Q^\prime}^i$.  Notice that the heavy quark number
operators $N_Q$ and $N_{Q^\prime}$ are now to be regarded as heavy quark
$1$-body operators
since one is not restricting to baryons with fixed numbers of $Q$ and
$Q^\prime$ quarks.  (Because of the constraint $N_c = N_\ell + N_Q +
N_{Q^\prime}$, it is not necessary to introduce $N_\ell$ as an additional
$1$-body operator as well.)  The coefficients of the expansion~(\ref{nexp})
contain implicit $N_Q/N_c$ and $N_{Q^\prime}/N_c$ dependence which appear as
operators in the generalized expansion.

\section{Masses of Heavy Baryons with Fixed Strangeness}

In this section, the masses of baryons containing heavy quarks are
analyzed in the combined $1/m_Q$ and $1/N_c$ expansion for the special case of
fixed strangeness.  It is useful to study this special case because it
illustrates the structure of the combined expansion without the complication of
flavor $SU(3)$ breaking.  The analysis here depends only on the presence of
isospin flavor symmetry.  The general $SU(3)$ mass analysis is given in Secs.~V
and~VI. 
  
The baryons containing a single heavy quark $Q$ with zero strangeness 
are the $\Lambda_Q$, $\Sigma_Q$ and $\Sigma_Q^*$.  The HQET $1/m_Q$ expansion
of these masses is given by
\begin{eqnarray}\label{lss}
\Lambda_Q &&= m_Q + \bar \Lambda_T 
- {\lambda_{1T} \over {2m_Q}} + ... \nonumber\\
\Sigma_Q &&= m_Q + \bar \Lambda_S - {\lambda_{1S} \over {2m_Q}}
-{{4\lambda_{2}} \over {2m_Q}} + ...\\
\Sigma_Q^* &&= m_Q + \bar \Lambda_S - {\lambda_{1S} \over {2m_Q}}
+{{2\lambda_{2}} \over {2m_Q}} + ... \nonumber
\end{eqnarray}
to order $1/m_Q^2$ in the $1/m_Q$ expansion.  The leading term in this
expansion is exactly $m_Q$ to all orders in the $1/N_c$ expansion.  Each of the
subleading parameters $\bar \Lambda$, $\lambda_{1}$ and $\lambda_2$ has an
expansion in $1/N_c$,
\begin{eqnarray}\label{hqexp}
&&\bar \Lambda  = c_0 N_c \openone
+ c_2 {1 \over N_c} J_\ell^2,
\nonumber\\
&&{{\lambda_{1}} \over {2 m_Q} } =
c_0^\prime {1 \over {2 m_Q}} N_Q
+ c_2^\prime {1 \over N_c^2} {1 \over {2 m_Q}} N_Q J_\ell^2,  \\
&&-d_H {{\lambda_2} \over {2 m_Q} } = c_2^{\prime\prime} {2 \over {
N_c m_Q}} \left( J_\ell \cdot J_Q \right),  \nonumber
\end{eqnarray}
where the factor of 2 in the $\lambda_2$ expansion follows from
the factor of $4$ in the
definition of $d_H = -4 \left( J_\ell \cdot J_Q \right)$.
The unknown coefficients in Eq.~(\ref{hqexp})
are a function of $1/N_c$ and the QCD scale
$\Lambda_{\rm QCD}$.    
Each coefficient has an expansion in $1 / N_c$
beginning at order unity in the $1/N_c$ expansion.
(The coefficients do not depend on $1/m_Q$
since the HQET parameters are defined to occur at a definite order in the
$1/m_Q$ expansion.)
Without loss of generality, it is possible
to set $c_0 \equiv \Lambda$, and use this relation as the definition of a QCD
scale.  The remaining coefficients are now given by a dimensional power of
$\Lambda$ times a dimensionless function of $1/N_c$ beginning at order unity in
the $1/N_c$ expansion.  For example, the coefficient $c_2$ is proportional to
$\Lambda$, whereas the $1/m_Q$-suppressed operators have coefficients
proportional to $\Lambda^2$.

The numbers of operators of the $1/N_c$ expansions for $\bar \Lambda$, $\lambda_1$ and
$\lambda_2$ are in one-to-one correspondence with the numbers of
parameters $\bar \Lambda_T$, $\bar \Lambda_S$, $\lambda_{1T}$,
$\lambda_{1S}$ and $\lambda_2$ appearing in
Eq.~(\ref{lss}).  The $1/N_c$ expansion predicts that the $\bar \Lambda$
parameters of the $\bar 3$ and $6$ are equal to $N_c \Lambda$ with a splitting
of relative order $1/N_c^2$ compared to the leading contribution.  Similarly,
the $\lambda_1$ parameters of the $\bar 3$ and $6$ are order unity in the
$1/N_c$ expansion with a splitting of relative order $1/N_c^2$.  Finally, the
$\lambda_2$ parameter for baryons is order $1/N_c$.  The $1/N_c$ scaling of
$\lambda_2$ can be tested by comparing the meson and baryon hyperfine
splittings resulting from the heavy quark spin symmetry-violating
chromomagnetic operator.  The baryon mass splitting is only measured in the
charm system at present\footnote{Recent measurements of the $\Sigma_b^{(*)}$
masses by DELPHI are probably unreliable and will be ignored.}.  The
charm meson hyperfine mass splitting is given by
\begin{equation}
\left( D^* - D \right) ={ { 2 \lambda_2^{\rm meson}} \over m_c} +...,
\end{equation}
whereas the analogous charm baryon mass splitting is
\begin{equation}
\left( \Sigma_c^* - \Sigma_c \right) = {{3 \lambda_2^{\rm baryon}} \over m_c}+...
\end{equation}
The naive $1/N_c$ scaling predicted by the baryon $1/N_c$ expansion implies
\begin{equation}
\lambda_2^{\rm baryon} \sim {1 \over N_c} \lambda_2^{\rm meson}.
\end{equation}
The measured charm meson and baryon mass differences $\left( D^* - D \right) =
141$~MeV and $\left( \Sigma_c^* - \Sigma_c \right) =77.1 \pm 5 \pm 5$~MeV are
in remarkable agreement with this scaling.
The $(D^* -D)$ mass
difference determines the canonical heavy quark symmetry
suppression factor $\Lambda_{\rm QCD}^2/m_c$,
\begin{equation}
{\lambda_2^{\rm meson} \over m_c} \sim {\Lambda_{\rm QCD}^2 \over m_c} 
= \frac 1 2 (D^* - D) .
\end{equation}

The $\lambda_1$ parameters for baryons and mesons also can be related.
The $1/N_c$ expansion implies
\begin{equation}\label{lone} 
\lambda_1^{\rm baryon} \sim \lambda_1^{\rm meson}\ .
\end{equation}
Eq.~(\ref{lone}) could be used to estimate the magnitude of 
$\lambda_1^{\rm baryon}$ if $\lambda_1^{\rm meson}$ is known accurately.
The most recent extraction of $\lambda_1^{\rm meson}$\cite{gklw} is too
uncertain, however, so the naive estimate 
\begin{equation}\label{loneb}
{\lambda_1^{\rm baryon} \over {2 m_Q} } 
\sim {{\Lambda_{\rm QCD}^2} \over {2 m_Q}},
\end{equation} 
is used in Eq.~(\ref{hqexp}).

We now proceed to analyze the masses of baryons containing heavy quarks using
the combined $1/m_Q$ and $1/N_c$ operator expansion developed in the previous
section.  We consider several different mass expansions.

Let us first consider the operator expansion for strangeness zero
baryons containing a single flavor of heavy quark $Q$.  The three baryons
masses $\Lambda_Q$, $\Sigma_Q$ and $\Sigma_Q^*$ are parametrized by the
three operators\cite{j},  
\begin{eqnarray}\label{oneq}
M = \left( m_Q + N_c \Lambda +... \right)\openone 
+ {1 \over N_c} J_\ell^2 + {2 \over {N_c m_Q}} \left( J_\ell \cdot J_Q \right),
\end{eqnarray}
where the leading $N_c$ and $m_Q$ dependence of each operator appears
explicitly.  The leading operator $\openone$ has a coefficient given exactly
by $m_Q + N_c \Lambda$ upto a correction of order $1/m_Q$ in the combined
$1/m_Q$ and $1/N_c$ expansion. 
It is to be understood that every other operator
is accompanied by a coefficient
\begin{equation}\label{coeff}
c\left(\Lambda,{1\over m_Q},{1 \over N_c}\right)
\end{equation}
with an expansion in $1/m_Q$ and $1/N_c$ beginning at order unity.  
For example, expanding to order $1/m_Q^2$ in the $1/m_Q$ expansion, there is an
order $(1/ 2 m_Q)$ contribution to the operator $\openone$ from the $\lambda_1$
contribution proportional to $c_0^\prime$ in Eq.~(\ref{hqexp}) 
and an order $(1/N_c^2)(1/ 2 m_Q)$ 
contribution to the operator $J_\ell^2$ from the $\lambda_1$ contribution
proportional to $c_2^\prime$.  Each of the operators in Eq.~(\ref{oneq})
contributes to one specific linear combination of the three baryon masses.
The operators and their corresponding mass combinations 
are tabulated in Table~I, together with the operator matrix element and the
leading $1/m_Q$ and $1/N_c$ dependence for each mass splitting.  The operator
matrix elements depend implicitly on $N_\ell=2$ and $N_Q=1$.  Up to unknown
coefficients of order unity, the combined $1/m_Q$ and $1/N_c$ expansion
predicts that the mass combinations $\Lambda_Q$, $\frac 1 3 \left(
\Sigma_Q + 2 \Sigma_Q^* \right) - \Lambda_Q$, and 
$\left(\Sigma_Q^* - \Sigma_Q\right)$
satisfy a hierarchy given by $\left(m_Q + N_c \Lambda\right)$, $2 \Lambda/ N_c$
and $\frac 3 2 (2\Lambda^2/N_c m_Q)$.  For $Q=c$, the
experimental values of the mass splittings are $2285.0 \pm 0.6$~MeV,
$219\pm 3 \pm 3$~MeV and $77\pm5\pm5$~MeV, which compares very favorably with
the theoretical hierarchy for canonical values of 
$m_c$ and $\Lambda$\footnote{The parameter values $m_c= 1450$~MeV,
$\Lambda_{\rm QCD} = \Lambda =310$~MeV, $\Lambda_{\rm QCD}^2/m_c = 66$~MeV,
$\Lambda_{\rm QCD}/m_c = 0.21$, $m_b = 4757$~MeV and $m_c/m_b =0.3$ are used
throughout this paper.  These parameter values follow from the
charm baryon mass analysis of this work, and are consistent with canonical
values used in the heavy quark literature.}.  
(Note that the experimental values and errors of
the hyperfine splittings are dominated by the uncertainty in the mass
measurement $\Sigma_c^* = 2530 \pm 5 \pm 5$~MeV.)  
   
For two flavors of heavy quark $Q=c$ and $Q^\prime=b$, the previous analysis of
the baryon mass spectrum holds for each flavor of heavy quark separately.
In the presence of heavy quark flavor symmetry,
a combined analysis of the six heavy baryon masses using heavy quark
spin-flavor operators is possible.  This mass expansion for baryons containing
a single heavy quark of flavor $Q$ or $Q^\prime$ is given by
\begin{eqnarray}\label{twoq}
M &&= \left[ \frac 1 2 \left( m_Q + m_{Q^\prime}\right) 
+ N_c \Lambda + ... \right] \openone 
+ I^3_h \left( m_Q - m_{Q^\prime}\right) \nonumber\\
&&+ {1 \over N_c} J_\ell^2 
+ {1 \over N_c^2} I_h^3 (J_\ell)^2 
\left( {1 \over {2 m_Q}} - {1 \over {2 m_{Q^\prime}}} \right) \nonumber\\
&&+  {1 \over {N_c}} \left( J_\ell \cdot J_h \right)
\left( {1 \over m_Q} + {1 \over m_{Q^\prime}} \right)\\
&&+ {1 \over {N_c }} J_\ell^i G_h^{i3} 
\left({2 \over m_Q} - {2 \over m_{Q^\prime}} \right),
\nonumber
\end{eqnarray}
where leading $1/m_Q$ and $1/N_c$ dependences are given
explicitly and unknown coefficients of the form Eq.~(\ref{coeff}) are
understood to accompany each operator [with the exception of the order
$m_Q$ and $N_c\Lambda$ terms given explicitly on the first line of
Eq.~(\ref{twoq})].  The important new operators of the heavy quark spin-flavor
analysis are the heavy quark flavor-violating operators which do not violate
heavy quark spin symmetry:  $I_h^3$ and $I_h^3 \left( J_\ell \right)^2$.
The most interesting operator is $I_h^3 \left( J_\ell \right)^2$.
The leading order $J_\ell^2$ hyperfine splitting for heavy baryons is a heavy
quark flavor-independent splitting of order $1/N_c$.  The $I_h^3 \left( J_\ell
\right)^2$ operator represents heavy quark flavor symmetry-breaking in the
$J_\ell^2$ hyperfine splittings.  This heavy flavor symmetry-breaking hyperfine
mass splitting is order $1/N_c^2$ times $\left( 1/2m_Q - 1/2m_{Q^\prime}
\right)$, and is sensitive to the subleading $(1/N_c^2)(1/ 2m_Q)$ contribution
to the $J_\ell^2$ hyperfine splitting.
Similar remarks hold for the operator
$I_h^3$, which represents heavy quark flavor symmetry breaking in the singlet
$\openone$ mass, 
except that this case is less interesting since there is a leading
contribution equal to $\left(m_Q - m_{Q^\prime}\right)$ which dominates the
subleading contribution proportional to $\left(1/2m_Q -
1/2m_{Q^\prime}\right)$. 

The mass combinations corresponding to the operators $I_h^3$ and $I_h^3
\left(J_\ell \right)^2$ are tabulated in Table~II.  
Both operators probe heavy flavor violation in $\lambda_1$.
The latter operator is sensitive to the
the $\lambda_1$ contribution
proportional to $c_2^\prime$, while the former
operator depends on the $\lambda_1$ operator proportional to $c_0^\prime$ only
at subleading order.  
The $\Sigma_b^{(*)}$ masses are not reliably measured at present, so the $I_h^3
\left( J_\ell \right)^2$ splitting cannot be evaluated.  Instead, the mass
relation
\begin{equation}\label{cbsig}
\frac 1 3 \left( \Sigma_b + 2 \Sigma_b^* \right) - \Lambda_b =
\frac 1 3 \left( \Sigma_c + 2 \Sigma_c^* \right) - \Lambda_c
\end{equation} 
can be used to predict the bottom baryon $J_\ell^2$ hyperfine splitting to
a theoretical accuracy of order $2 /N_c^2$ times $\Lambda^2
\left( 1/ 2m_c - 1/ 2m_b
\right)$, which is about $5$~MeV.  Thus, the mass splitting 
$\left[ \frac 1 3 \left( \Sigma_b + 2 \Sigma_b^* \right) - \Lambda_b \right]$
is predicted to be equal to $219 \pm 7$~MeV using Eq.~(\ref{cbsig}).
The measured $\left(\Lambda_c^+ - \Lambda_b^0\right)$
splitting of $-3338 \pm 5 \pm 4$~MeV gives a measure of $\left(m_c -
m_b\right)$, upto corrections of order $\Lambda^2\left( 1/ 2m_c - 1/ 2m_b
\right)$ which is about $23$~MeV. 

The remaining operators in the expansion~(\ref{twoq}) correpond to sum and
difference combinations of $Q$ and $Q^\prime$ splittings.  There is no
advantage 
to studying these combinations rather than the $Q$ and $Q^\prime$ splittings
separately, since there is no cancellation of leading order contributions in
these terms.  For example, there is no reason to study $J_\ell \cdot J_h$
and $J_\ell^i G_h^{i3}$ rather than $J_\ell \cdot J_Q$ and $J_\ell \cdot
J_{Q^\prime}$, which are order $2/N_c m_Q$ and $2/N_c m_{Q^\prime}$, respectively.  

Additional information about the heavy quark baryon mass spectrum can be
obtained by generalizing the expansion to encompass
baryons with differing number of heavy quarks.  The
strangeness zero baryons for one flavor of heavy quark $Q$
consist of the $N_Q=0$ baryons $N$ and $\Delta$;
the $N_Q=1$ baryons $\Lambda_Q$, $\Sigma_Q$, and $\Sigma_Q^*$; the $N_Q=2$
baryons $\Xi_{QQ}$ and $\Xi_{QQ}^*$; and the $N_Q=3$ baryon $\Omega_{QQQ}^*$.
The $1/N_c$ operator basis for strangeness zero baryons with $N_Q = 0,1,2,3$
heavy quarks consists of the eight operators $\openone$, $N_Q$,
$J_\ell^2$, $\left(J_\ell \cdot J_Q\right)$, $N_Q J_\ell^2$, $N_Q^2$, 
$N_Q \left( J_\ell \cdot J_Q \right)$, and $N_Q^3$. Notice that $N_Q$ is to be
regarded as a $1$-body heavy quark operator along with $J_Q$ when considering
baryons with differing numbers of heavy quarks. 
Although it is interesting to study this
expansion for $N_Q=0,1,2,3$ baryons, the principal interest of this work will
be to relate mass splittings of baryons containing a single heavy quark
to the well-measured mass splittings of baryons containing no heavy quarks,
namely baryons in the spin-$\frac 1 2$ octet and spin-$\frac 3 2$ decuplet.  
Thus, we will restrict the operator expansion to baryons with $N_Q=0$ and $1$.
The operator
expansion for strangeness zero baryons containing no heavy quark and baryons
containing a single heavy quark of flavor $Q$ is given by  
\begin{eqnarray}\label{mq}
M &&= N_Q \left( m_Q + ...\right) + N_c \Lambda \openone
+ {1 \over N_c} J_\ell^2 \nonumber\\
&&+ {2 \over {N_c m_Q}} \left( J_\ell \cdot J_Q \right)
+ {1 \over N_c^2} N_Q J_\ell^2,
\end{eqnarray}
where leading $1/m_Q$ and $1/N_c$ dependences are given explicitly and unknown
coefficients of the form Eq.~(\ref{coeff}) are understood to accompany each
operator with the exception of the order $m_Q$ and $N_c\Lambda$ terms.    
Since one is only considering baryons containing upto one heavy
quark, each operator in the expansion has at most one heavy quark $1$-body 
operator $N_Q$ or $J_Q$.   
Comparing
with Eq.~(\ref{oneq}), there are two new operators in the present
expansion: $N_Q$, $N_Q J_\ell^2$.  These operators reduce to $\openone$ and
$J_\ell^2$ for baryons with fixed $N_Q=1$, reproducing expansion~(\ref{oneq}).
The most interesting operator is $N_Q J_\ell^2$ which represents the heavy
quark number-dependent contribution to the $J_\ell^2$ hyperfine splittings.
This heavy quark number-dependent mass splitting is order 
$(1/(N_c^2)$.  The $N_Q$ operator
is less interesting since it corresponds to the leading $N_Q$-dependent mass
splitting which is equal to $m_Q$.  There is a subleading order unity
contribution to this splitting.  

The mass combinations corresponding to $N_Q$ and $N_Q J_\ell^2$
are tabulated in Table~III.    
The $\left[\Lambda_Q - \frac 1 4 \left( 5 N - \Delta \right)\right]$ splitting
of $1419.2 \pm 0.7$ and $4757.2 \pm 6.4$ for $Q=c$ and $b$, respectively, give
a measure of $m_c$ and $m_b$, upto corrections of order $\Lambda$.  The $N_Q
J_\ell^2$ mass relation,
\begin{equation}\label{hypq}
\frac 1 3 \left( \Sigma_Q + 2 \Sigma_Q^* \right)- \Lambda_Q 
= \frac 2 3 \left( \Delta - N \right),
\end{equation}
is predicted to be satisfied to order $(1/N_c^2)$.
Relation~(\ref{hypq}) has 
been obtained previously in the context of the Skyrme model where heavy baryons
appear as bound states of Skyrmions and heavy mesons\cite{ck,skyrme}.  The present derivation
is model-independent, and provides a prediction for the accuracy to which the
relation is satisfied.  The theoretically predicted accuracy can be tested in
the charm baryon sector where the relevant masses are measured.   
The $(\Delta -N )$ splitting of $292 \pm 1$~MeV
and the $\left[\frac 1 3 (\Sigma_c + 2 \Sigma_c^* ) - \Lambda_c\right]$
splitting of $219 \pm 4.6$~MeV are both $J_\ell^2/N_c$ splittings with
$J_\ell^2$ matrix elements of $3$ and $2$, respectively.
The mass relation~(\ref{hypq}) is
satisfied to $24 \pm 5$~MeV, to be compared with the predicted
violation of $(1/N_c^2)\Lambda$ times the matrix
element $\langle N_Q J_\ell^2 \rangle=2$, which is about $69$~MeV.  
(Alternatively, the violation should
be suppressed by a factor of $(1/N_c)$ compared to the charm
baryon $J_\ell^2$ splitting of $219$~MeV.)
Relation~(\ref{hypq}) can be used to 
predict the bottom baryon $J_\ell^2$ hyperfine splitting 
$\left[\frac 1 3 (\Sigma_b + 2 \Sigma_b^* ) - \Lambda_b\right]$ to an accuracy
of $(1/N_c^2)$ as well.  The theoretical prediction is
\begin{equation}\label{hypb}
\frac 1 3 (\Sigma_b + 2 \Sigma_b^* ) - \Lambda_b = 195 \pm 69~{\rm MeV},
\end{equation}
which is consistent with, but much less accurate than, the prediction 
made from the
previous expansion using the $I_h^3 \left(J_\ell\right)^2$ mass combination.

It is instructive to generalize the expansion~(\ref{mq}) to two heavy quark
flavors $Q$ and $Q^\prime$.  The mass expansion for strangeness zero baryons
containing zero or one heavy quark of flavor $Q$ or $Q^\prime$ is given by  
\begin{eqnarray}
M &&= N_Q \left( m_Q + ... \right) 
+ N_{Q^\prime} \left( m_{Q^\prime} + ...\right)
+ N_c \Lambda \openone \nonumber\\
&&+ {1 \over N_c} J_\ell^2 
+ {1 \over N_c^2} N_Q J_\ell^2
+ {1 \over N_c^2} N_{Q^\prime} J_\ell^2 \\
&&+ {2 \over {N_c m_Q}} \left( J_\ell \cdot J_Q \right)
+ {2 \over {N_c m_{Q^\prime}}} \left( J_\ell \cdot J_{Q^\prime} \right),
\nonumber
\end{eqnarray}
where leading $1/m_Q$ and $1/N_c$ dependences are given explicitly and unknown
coefficients are understood to accompany all operators except the $m_Q$ and
$N_c \Lambda$ terms.  This expansion can be rewritten in terms of heavy quark
spin-flavor operators as follows,
\begin{eqnarray}
M &&= \frac 1 2 N_h \left[\left( m_Q + m_{Q^\prime} \right)+ ... \right]
+ I^3_h \left( m_Q - m_{Q^\prime}\right) + N_c \Lambda \openone \nonumber\\
&&+ {1 \over N_c} J_\ell^2 
+ {1 \over N_c^2} N_h J_\ell^2 \nonumber\\
&&+ {1 \over N_c^2} I_h^3 (J_\ell)^2 
\left( {1 \over  {2m_Q}} - {1 \over  {2 m_{Q^\prime}}} \right) \\
&&+ {1 \over {N_c}} \left( J_\ell \cdot J_h \right)
\left( {1 \over m_Q} + {1 \over m_{Q^\prime}} \right) \nonumber\\
&&+ {1 \over {N_c }} J_\ell^i G_h^{i3} 
\left({2 \over m_Q} - {2 \over m_{Q^\prime}} \right)
\nonumber
\end{eqnarray}
where $N_h = N_Q + N_{Q^\prime}$.  The interesting suppressions of the heavy
quark flavor symmetry-violating and heavy quark number-dependent $J_\ell^2$
splittings discussed in the preceeding paragraphs are manifest.

The above analysis which applies to $S=0$ baryons in the isospin limit
also can be performed in the $S=-1$ sector.  These mass combinations are
provided in Tables~I,~II and~III as well.  The charm baryon masses $\Xi_c$ and
$\Xi_c^*$ are well-measured, but the lone $\Xi^\prime_c$ mass measurement is
unpublished and possibly unreliable.  A precise measurement of the
$\Xi_c^\prime$ is likely at CLEO in the not too distant future.  A prediction
of the $\Xi_c^\prime$ mass using the $N_Q J_\ell^2$ mass relation
\begin{equation}\label{xhyp}
\frac 1 3 \left( \Xi_Q^\prime + 2 \Xi_Q^* \right) - \Xi_Q =
\frac 2 3 \left[ \Sigma^* 
- \frac 1 4 \left( 3 \Sigma + \Lambda \right) \right]
\end{equation}
is possible, but not terribly accurate.  Relation~(\ref{xhyp}) is predicted to
be satisfied to an accuracy of order $\left(2 \Lambda/N_c^2  \right)$,
which is about $69$~MeV.  
The 
$\left[\Sigma^* - \frac 1 4 \left( 3 \Sigma + \Lambda\right)\right]$ mass
splitting is $210.8 \pm 0.4$~MeV, so Eq.~(\ref{xhyp}) predicts $\left[\frac 1 3
\left( \Xi^\prime_c + 2 \Xi_c^* \right) - \Xi_c \right] = 140.5 \pm 69$~MeV.
The $\Xi_c \equiv \frac 1 2 \left( \Xi_c^+ + \Xi_c^0 \right)$
mass is $2467.7 \pm 1.2$~MeV and the analogous $\Xi_c^*$ mass is $2644.0 \pm
1.6$~MeV, so the 
$\Xi_c^\prime$ prediction is not very useful.  A much more
precise prediction of the $\Xi_c^\prime$ mass is made in Sect.~V.
The bottom baryon $S=-1$ $J_\ell^2$ splitting also is predicted by
relation~(\ref{xhyp}) to the same theoretical accuracy.
Once the $\Xi_c^\prime$ mass is measured, the bottom baryon $J_\ell^2$
splitting can be accurately predicted from the charm baryon mass splitting.  

The results of this section follow from the combined $1/m_Q$ and $1/N_c$
expansion using only isospin symmetry of the $u$ and $d$ quarks.  The $SU(3)$
analysis including $SU(3)$ flavor breaking is provided in the next section.
The $SU(3)$ analysis leads to more precise predictions than the isospin
analysis.  The isospin symmetry predictions of this section are consistent with
the mass predictions of Sec.~V.
The isospin-breaking analysis is given in Sec.~VI.

\section{$SU(3)$ Analysis of Heavy Baryon Masses}

The $SU(3)$ analysis of the masses of baryons containing a single heavy quark
in the combined $1/m_Q$ and $1/N_c$ expansion is a straightforward
generalization of the isospin
analysis of the previous section.  The lowest-lying
baryons containing a single heavy quark are the spin-$\frac 1 2$ $\bar 3$
which consists of the isosinglet $\Lambda_Q$ and the isodoublet $\Xi_Q$; the
spin-$\frac 1 2$ $6$ which consists of the isotriplet $\Sigma_Q$, the
isodoublet $\Xi_Q^\prime$, and the isosinglet $\Omega_Q$; and the spin-$\frac 3
2$ $6$ which consists of the isotriplet $\Sigma_Q^*$, the isodoublet $\Xi_Q^*$
and the isosinglet $\Omega_Q^*$.  Heavy baryon masses will be denoted by 
their particle labels.  
Each of the $I=0$ masses refers
to the average mass of the isomultiplet.  For example,
$\Xi_c = \frac 1 2 \left( \Xi_c^+ + \Xi^0 \right)$, etc.  
In addition to the eight $I=0$ heavy baryon
masses, there is an $I=0$ off-diagonal mass $\bar \Xi_Q \Xi_Q^\prime = \bar
\Xi_Q^\prime \Xi_Q$ between the spin-$\frac 1 2$ isodoublets,
which will be referred to as $\Xi_Q \Xi_Q^\prime$ throughout
the rest of the paper.
The $I=0$ off-diagonal mass is defined to be the average off-diagonal mass of
the spin-$\frac 1 2$ $S=-1$ isodoublets, 
\begin{eqnarray}
\Xi_c \Xi_c^\prime &&= \frac 1 2 \left( \Xi_c^+ \Xi_c^{\prime +} + \Xi_c^0
\Xi_c^{0 \prime} \right), \nonumber\\
\Xi_b \Xi_b^\prime &&= \frac 1 2 \left( \Xi_b^0 \Xi_b^{\prime 0} + \Xi_b^-
\Xi_b^{- \prime} \right).
\end{eqnarray}

Many of the charm
baryon masses are now measured.  The measured $I=0$ masses are
\begin{eqnarray}
&&\Lambda_c = 2285.0 \pm 0.6~{\rm MeV} \cite{pdg}, \nonumber\\
&&\Xi_c = 2467.7 \pm 1.2~{\rm MeV}\cite{pdg}, \nonumber\\
&&\Sigma_c = 2452.9 \pm 0.6~{\rm MeV}\cite{pdg,e687,cleoi}, \\
&&\Omega_c^0 = 2704 \pm 4~{\rm MeV}\cite{pdg}, \nonumber\\
&&\Xi_c^* = 2644.0 \pm 1.6~{\rm MeV}\cite{cleoii}, \nonumber
\end{eqnarray}  
The masses $\Xi_c^\prime \sim 2560$~MeV\cite{wa89} and $\Sigma_c^* = 2530 \pm 5 \pm 5$~MeV\cite{pdg}
are measured, but are less reliable or precise.  The $\Omega_c^*$ mass and the
off-diagonal mass $\Xi_c \Xi_c^\prime$ are unmeasured.
At present, only the
bottom baryon 
\begin{equation}
\Lambda_b = 5623 \pm 5 \pm 4~{\rm MeV} \cite{cdf}
\end{equation}
is accurately measured.  As stated earlier, the reported $\Sigma_b^{(*)}$
measurements by DELPHI are not used.   

The operator analysis including $SU(3)$ flavor breaking
decomposes into $SU(3)$ flavor singlet, octet and 
$27$\footnote{The analytic flavor
dependence of the flavor-27 expansion is given here.  The non-analytic
contribution of order $\epsilon^{3/2}$ is proportional to $\left(\frac 1 3
m_\pi^2 - \frac 4 3 m_K^2 + m_\eta^2\right)$ which is numerically very small,
and does not dominate the $\epsilon^2$ counterterm\cite{jmw,jcpt}. 
The order $\epsilon^{3/2}$ contribution is computed in chiral perturbation
theory in Ref.~\cite{savage}.}
expansions:
\begin{equation}\label{msot}
M = M^{(1)} + \epsilon M^{(8)} + \epsilon^2 M^{(27)},
\end{equation}
where $\epsilon$ is an $SU(3)$-violating parameter whose magnitude is governed
by the quark mass difference $\left( m_s - \hat m \right)$ divided by the
chiral symmetry breaking scale $\Lambda_\chi$.  The symmetry breaking parameter
$\epsilon \sim 0.25$.
The flavor singlet, octet and $27$ mass expansions are given by
\begin{eqnarray}\label{sot}
&&M^{(1)} = \left( m_Q + N_c \Lambda +... \right)\openone 
+ {1 \over N_c} J_\ell^2 
+ {1 \over N_c}{2 \over {m_Q}} \left( J_\ell \cdot J_Q \right),
\nonumber\\
&&M^{(8)} = T^8 + {1 \over N_c} J_\ell^i G^{i8} + {1 \over N_c}
{2 \over { m_Q}} J_Q^i G^{i8} \nonumber\\ 
&&\qquad + {1 \over N_c^2}{2 \over { m_Q}} \left( J_\ell \cdot J_Q
\right) T^8, \\ 
&&M^{(27)} = {1 \over N_c} \{ T^8, T^8 \} + {1 \over N_c^2}{2 \over { m_Q}}
J_Q^i \{ T^8, G^{i8} \}, \nonumber
\end{eqnarray}
where leading $1/m_Q$ and $1/N_c$ dependences are given explicitly and unknown
coefficients are understood to accompany all operators (except for the leading
terms of the $\openone$ operator).  The operators appearing in the flavor-$27$
expansion require flavor singlet and octet subtractions: these subtractions are
implied, but not given explicitly throughout this paper.   The relevant
flavor-$27$ projection operator for
an operator ${\cal O}^{ab}$ with two symmetric flavor indices $(ab)$ is
\begin{equation}
\delta^{a8} \delta^{b8} - \frac 1 8 \delta^{ab} - \frac 3 5 d^{ab8} d^{888} .
\end{equation}  
The operators and their corresponding mass
combinations are tabulated in Table~IV, together with the $SU(3)$ and $J_Q$
representations of the operator, the operator matrix
element, and the leading $1/m_Q$, $1/N_c$ and $SU(3)$ flavor-breaking
factors.  Table~V combines these factors to predict a hierarchy of mass
combinations.  The numerical values of the theoretical hierarchy are evaluated
for canonical values of parameters $m_Q$, $\Lambda$, $(\epsilon \Lambda_\chi)$
and $(\epsilon^2 \Lambda_\chi)$\footnote{Canonical values for the flavor octet
and flavor-$27$ mass splittings are $\epsilon \Lambda_\chi = 225$~MeV and
$\left( \epsilon^2 \Lambda_\chi \right)= 8.8$~MeV.  These values are obtained
from the octet and decuplet baryon masses.  Note that the flavor-$27$ mass
parameter is considerably smaller than $\epsilon$ times the octet mass
parameter, and may be underestimated.} in the columns labeled $Q=c$ and $Q=b$.
It is not possible to evaluate seven of the nine mass combinations using the
measured charm baryon masses without knowledge of $\Omega_c^*$ and $\Xi_c
\Xi_c^\prime$.  The measured mass combinations $\frac 1 3 (\Lambda_c + 2 \Xi_c
)$ and $(\Lambda_c - \Xi_c )$ which appear in column four were used to 
determine the parameters $(m_c + N_c \Lambda)$ and $(\epsilon \Lambda_\chi)$,
and so do not test the theoretical hierarchy. 

Table~V predicts a hierarchy of mass relations obtained 
by successively neglecting operators
in the mass expansion~(\ref{msot}).  The most accurate mass relation is the
flavor-$27$ heavy quark spin-violating relation
\begin{equation}\label{mreli}
\frac 1 4 \left[ \left( \Sigma_Q^* - \Sigma_Q \right)
- 2 \left( \Xi^*_Q - \Xi^\prime_Q \right)
+ \left( \Omega^*_Q - \Omega_Q \right) \right]=0,
\end{equation}
which is satisfied up to a correction of
order $\frac 9 {8} (1/N_c^2) (\Lambda^2/m_Q)(\epsilon^2 \Lambda_\chi)$, 
which is about $0.23$ MeV for $Q=c$ and $0.07$ for $Q=b$.  Thus,
Eq.~(\ref{mreli}) is essentially an exact relation for both the charm and the
bottom baryons.  The physical content of this relation is that the three
chromomagnetic mass splittings of the sextet, 
$\left(\Sigma_Q^* - \Sigma_Q\right)$,
$\left(\Xi_Q^* -\Xi^\prime_Q\right)$ and $\left(\Omega^*_Q - \Omega_Q\right)$,
contain only a singlet contribution and a contribution which is linear in
strangeness.  Differences of the splittings are equal,
\begin{eqnarray}\label{chresr}
&&\left(\Sigma_Q^* - \Sigma_Q\right)-\left(\Xi_Q^* -\Xi^\prime_Q\right)
\nonumber\\
&&=\left(\Xi_Q^* -\Xi^\prime_Q\right)-\left(\Omega^*_Q - \Omega_Q\right)\ .
\end{eqnarray}
Eq.~(\ref{chresr}) is the equal spacing rule of Savage \cite{savage} for the
chromomagnetic mass splittings of the sextet.

The next most accurate relation is the flavor-$27$ mass relation,
\begin{equation}\label{jhypi}
\frac 1 6 \left[ \left( \Sigma_Q +2 \Sigma_Q^* \right) -
2  \left( \Xi^\prime_Q +2 \Xi^*_Q \right) + \left( \Omega_Q +2 \Omega_Q^*
\right) \right] =0,
\end{equation}
which is expected to be satisfied to order $\frac 3 2 (1/N_c)(\epsilon^2
\Lambda_\chi)$, or about $4.4$~MeV for $Q=c,b$.  This mass relation
respects heavy quark spin symmetry.  It implies that the three   
spin-averaged masses of the sextet, $\frac 1 3 \left(
\Sigma_Q + 2 \Sigma_Q^* \right)$, $\frac 1 3 \left( \Xi_Q^\prime + 2 \Xi_Q^*
\right)$ and $\frac 1 3 \left( \Omega_Q + 2 \Omega_Q^* \right)$, receive a
singlet contribution and a contribution linear in strangeness.  Differences of
the splittings are predicted to be equal
\begin{eqnarray}\label{saesr}
&&\frac 1 3 \left( \Xi_Q^\prime + 2 \Xi_Q^*\right)-
\frac 1 3 \left(\Sigma_Q + 2 \Sigma_Q^* \right) \nonumber\\
&&=\frac 1 3 \left( \Omega_Q + 2 \Omega_Q^* \right)
-\frac 1 3 \left( \Xi_Q^\prime + 2 \Xi_Q^* \right), 
\end{eqnarray}    
at the $8.8$~MeV level.  Eq.~(\ref{saesr}) is an equal spacing rule 
for the spin-averaged masses of the sextet.

The two flavor-$27$ mass relations Eqs.~(\ref{mreli}) and~(\ref{jhypi})
together imply the vanishing of the spin-$\frac 1 2$ and spin-$\frac 3 2$
sextet flavor-$27$ combinations
\begin{eqnarray}\label{f27}
&&\Sigma_Q -2 \Xi_Q^\prime + \Omega_Q=0, \nonumber\\
&&\Sigma_Q^* -2 \Xi_Q^* + \Omega_Q^*=0\ ,
\end{eqnarray}
to a precision of about $8.8$~MeV.  The first of these equations was obtained
previously by Savage \cite{savage}.

There are two additional mass relations 
\begin{eqnarray}\label{mrelii}
&&\frac 1 6 \left[ 3 \left( \Sigma_Q^* - \Sigma_Q \right)
- \left( \Xi^*_Q - \Xi^\prime_Q \right)
-2 \left( \Omega^*_Q - \Omega_Q \right)\right] \nonumber\\
&&\qquad\qquad-{5 \over {2\sqrt{3}}} \Xi_Q \Xi_Q^\prime =0, \\
&&\qquad\qquad\Xi_Q \Xi_Q^\prime = 0, \nonumber
\end{eqnarray}
satisfied to orders 
$\frac {1} {2 \sqrt{3}}\frac {15} 4 (2 /N_c^2) (\Lambda / m_Q)(\epsilon
\Lambda_\chi)$  and
$\frac 3 4 (1/N_c)(\Lambda /m_Q)(\epsilon \Lambda_\chi)$, respectively,
which are numerically comparable: about $11.5$~MeV for $Q=c$ and $3.5$ for
$Q=b$.  The second relation gives an order of magnitude estimate of the $I=0$
off-diagonal mass $\Xi_Q \Xi_Q^\prime$, while the first relation implies
a determination of this off-diagonal mass in terms of the sextet masses.
The two relations together yield the mass relation
\begin{equation}\label{mreliv}
\frac 1 6 \left[ 3 \left( \Sigma_Q^* - \Sigma_Q \right)
- \left( \Xi^*_Q - \Xi^\prime_Q \right)
-2 \left( \Omega^*_Q - \Omega_Q \right)\right]=0,
\end{equation}
which is predicted to hold to $16.6 \pm 11.5$~MeV for $Q=c$ and to 
$5.1 \pm 3.5$~MeV for $Q=b$.
This relation combined with the flavor-$27$ relation~(\ref{mreli}) implies
that the chromomagnetic mass splittings of the sextet are equal,
\begin{equation}\label{esr}
\left(\Sigma_Q^* - \Sigma_Q\right) = \left(\Xi_Q^* - \Xi_Q^\prime\right) 
= \left(\Omega_Q^* - \Omega_Q\right).
\end{equation}
Eq.~(\ref{esr}) was obtained previously by Savage \cite{savage}.
The combined $1/m_Q$, $1/N_c$ and $SU(3)$ flavor-breaking expansion predicts
that this equation need not be very accurate.  The two equalities
in Eq.~(\ref{esr}) are predicted to be satisfied only to 
$19.9 \pm 13.8$~MeV for
$Q=c$ and to $6.1 \pm 4.2$~MeV for $Q=b$.  The equality $\left(\Sigma_Q^* -
\Sigma_Q\right)= \left(\Omega_Q^* - \Omega_Q\right)$ is predicted to be
satisfied only to $39.8 \pm 27.6$~MeV for $Q=c$ and 
$12.2 \pm 8.4$~MeV for $Q=b$.

The remaining mass combinations in Table~V
are not suppressed enough to yield useful mass relations.  

The two flavor-$27$
mass relations can be used to predict the poorly measured $\Xi_c^\prime$ mass
and the unmeasured $\Omega_c^*$ mass.  [The mass $\Sigma_c^* = 2530 \pm
7$~MeV is known to greater precision than the mass relation~(\ref{mreliv})
which would be needed to extract a third charm baryon mass.]  The masses
$\Xi_c^\prime = 2578.5 \pm 4.8$~MeV and $\Omega_c^*=2758.0 \pm 11.7$~MeV are
obtained using the flavor-$27$ sextet mass relations~(\ref{f27}).  A different,
more precise, extraction of these masses will be obtained shortly.

The analysis of Sec.~IV showed that additional constraints on the heavy baryon
hierarchy follow from studying heavy quark flavor
symmetry violation and heavy quark number dependence in the heavy baryon mass
splittings which do not violate heavy quark spin symmetry.  For the $SU(3)$
analysis, these mass splittings correspond to the operators $\openone$,
$J_\ell^2$, $T^8$ $J_\ell^i G^{i8}$, and $\{T^8, T^8 \}$ times $I_h^3$ and
$N_Q$, respectively.  It is useful to study the mass combinations corresponding
to these operators since the leading order contribution of each $J_Q=0$ $SU(3)$
splitting cancels out of the corresponding heavy quark flavor-violating and
heavy quark number-dependent splitting.

The $I_h=1$
heavy quark flavor symmetry-violating mass splittings are tabulated in
Table~VI.  These mass combinations correspond to
splittings between heavy charm and bottom baryons, and are suppressed by a
relative factor of $1/N_c$ times $\left( 1/2m_c - 1/2m_b \right)$ in comparison
to the leading heavy quark flavor symmetric contribution to the heavy baryon
mass splitting [except for the $I_h^3 \openone$ splitting which is proportional
to $\left( m_c - m_b \right)$].  The most accurate $I_h=1$ mass relations in
Table~VI can be used to predict the bottom
baryon mass splittings from the corresponding charm baryon mass splittings.     

Heavy quark number-violating mass splittings between baryons containing a
single heavy quark $Q$ and baryons containing no heavy quarks are given in
Table~VII.  These mass combinations relate heavy baryon mass splittings to
splittings amongst the spin-$\frac 1 2$ octet and spin-$\frac 3 2$ decuplet
baryons.  The mass combinations are suppressed by a relative factor of 
$(1/N_c)$ in comparison to the leading contribution to the heavy
baryon mass splitting (except for the $N_Q \openone$ mass splitting which is
proportional to $m_Q$).
The $N_Q$ and $N_Q J_\ell^2$ mass combinations are the $SU(3)$
generalizations of the $SU(2)$ mass combinations studied in Sec.~IV.  The
linear combination of octet and decuplet masses appearing for these operators
are the $\openone$ and $J_\ell^2$ mass combinations for $N_Q=0$ baryons derived
in Ref.~\cite{jl}.  The derivation of the flavor octet and flavor-$27$ mass
combinations is more subtle.  There are five flavor octet ($J_Q=0$) operators
amongst the $N_Q=0$ and $N_Q=1$ baryons: $T^8$, $J_\ell^i G^{i8}$,
$\left(J_\ell\right)^2 T^8$, $N_Q T^8$ and $N_Q J_\ell^i G^{i8}$.  
It is important to note that the $4$-body operator $N_Q
\left(J_\ell\right)^2 T^8$ is a redundant operator.  The mass
combination corresponding to $N_Q J_\ell^i G^{i8}$ is the mass combination
with vanishing matrix elements of the remaining four operators.  Since
the matrix element of $\left(J_\ell\right)^2 T^8$ must vanish on this combination, the
linear combination of octet and decuplet masses appearing in the $N_Q J_\ell^i
G^{i8}$ mass combination is not simply the $J_\ell^i G^{i8}$ mass combination
for $N_Q=0$ baryons\cite{jl}.  Similar remarks apply for the flavor $27$ mass
combinations.  There are three flavor-$27$ ($J_Q=0$) operators amongst the
$N_Q=0$ and $N_Q=1$ baryons: $\{T^8, T^8 \}$, $\{ T^8 , J_\ell^i G^{i8} \}$ and
$N_Q \{ T^8, T^8 \}$.  The $4$-body operator $N_Q \{T^8, J_\ell^i G^{i8} \}$ is
a redundant operator.  The mass combination corresponding
to the $N_Q \{ T^8, T^8 \}$ operator is the mass combination with vanishing
matrix elements of the two other operators.

There are two accurate mass relations obtained by neglecting heavy quark
number-violating operators.
Neglect of the flavor-$27$ operator $N_Q \{T^8, T^8 \}$ yields the mass
relation
\begin{eqnarray}\label{nqmreli}
&&\frac 1 6 \left[ \left( \Sigma_Q + 2 \Sigma_Q^* \right) - 2
\left(\Xi_Q^\prime + 2 \Xi_Q^* \right) + \left( \Omega_Q + 2 \Omega_Q^* \right)
\right]= \nonumber\\
&& \frac 1 3 \left[\frac 1 4 \left( 2 N - \Sigma - 3 \Lambda + 2 \Xi \right) +
\frac 1 7 \left( 4 \Delta - 5 \Sigma^* - 2 \Xi^* + 3 \Omega \right) \right],
\end{eqnarray}     
which is exact up to corrections of order 
$(3/2N_c^2)(\epsilon^2 \Lambda_\chi)$.  
The estimate of this correction is $1.5$~MeV for $Q=c,b$. 
Eq.~(\ref{nqmreli}) is a factor of $(1 /N_c)$ more
accurate than relation Eq.~(\ref{jhypi}), which sets the heavy baryon mass
combination equal to zero.  The linear
combination of octet and decuplet masses on the right-hand side of the equation
equals $-4.43$~MeV with negligible error, which agrees with the
estimated accuracy of Eq.~(\ref{jhypi}). 
Neglect of the $N_Q J_\ell^i G^{i8}$ operator yields the mass relation
\begin{eqnarray}\label{iii}
&&\left\{-\frac 5 8
\left( \Lambda_Q - \Xi_Q \right) \right.\nonumber\\
&&\left.+ \frac 1 {24} \left[ 3 \left(\Sigma_Q + 2
\Sigma_Q^*\right) - \left(\Xi^\prime_Q + 2 \Xi_Q^* \right)
-2 \left(\Omega_Q + 2 \Omega_Q^*\right)\right]\right\} \nonumber\\
&&\quad =-\frac 1 {24} \left( 8 N - 9 \Sigma + 3 \Lambda -2 \Xi \right)
+ \frac 1 {12} \left( 2 \Delta - \Xi^* - \Omega \right),
\end{eqnarray}
which is satisfied upto a correction of order $\frac 1 {2 \sqrt{3}}\frac {15}
8(1/N_c^2)(\epsilon \Lambda_\chi)$, which is about
$13.5$~MeV for $Q=c,b$.  The linear combination of
octet and decuplet masses on the right-hand side of this equation equals
$42.89\pm 0.19$~MeV, so there was not a useful mass relation involving this
heavy baryon mass splitting in expansion~(5.5).   
Finally, a third mass relation following from the neglect of the $N_Q T^8$
operator provides a test of the theoretical hierarchy.  This mass relation
\begin{eqnarray}\label{jgrel}
&&\left( \Lambda_Q - \Xi_Q \right) = \frac 1 8 \left( 6 N - 3 \Sigma + \Lambda
- 4 \Xi \right) \nonumber\\
&&\qquad\qquad\qquad - \frac 1 {20} \left( 2 \Delta - \Xi^* - \Omega \right) ,
\end {eqnarray}   
is predicted to hold to order $\frac 3 {2 \sqrt{3}} (1/N_c)
(\epsilon \Lambda_\chi)$, which is about $65$~MeV for $Q=c,b$.  
Evaluation of the mass
difference of $(\Lambda_c - \Xi_c)$ and the right-hand side of
Eq.~(\ref{jgrel}) yields $43$~MeV, which is a factor 
of $0.6$ times the coefficient unity estimation.

Elimination of the mass $\frac 1 3 (\Omega_Q + 2 \Omega_Q^*)$ between
Eq.~(\ref{nqmreli}) and~(\ref{iii}) yields a prediction for the heavy baryon
mass splitting
\begin{eqnarray}
&&\left[ \frac 1 3 \left( \Sigma_Q + 2 \Sigma_Q^* \right) - \Lambda_Q \right]
-\left[ \frac 1 3 \left( \Xi^\prime_Q + 2 \Xi_Q^* \right) - \Xi_Q \right]
\end{eqnarray}
in terms of octet and decuplet baryon mass splittings. 

\subsection{Predictions}

The two new mass relations Eq.~(\ref{nqmreli}) and~(\ref{iii})
together with the most precise relation Eq.~(\ref{mreli}) can be used
to extract values
for the $\Sigma_c^*$, $\Xi^\prime_c$ and $\Omega_c^*$,
\begin{eqnarray}\label{sxo}
&&\Sigma_c^* = 2532.5 \pm 5.3~{\rm MeV}, \nonumber\\
&&\Xi_c^\prime = 2582.9 \pm 2.5~{\rm MeV}, \\
&&\Omega_c^* = 2746.6 \pm 6.1~{\rm MeV} . \nonumber
\end{eqnarray}
The predicted $\Sigma_c^*$ mass is consistent with the current PDG value of
$2530 \pm 5 \pm 5$~MeV, while the predicted $\Xi_c^\prime$ mass is significantly larger than
the unpublished WA-89 measurement of $\sim 2560$~MeV.  All of the other charm
baryon masses are accurately measured, so the predicted masses for
$\Sigma_c^*$, $\Xi_c^\prime$ and $\Omega_c^*$ together with the
measured charm baryon masses can be used
to evaluate the mass combinations of the combined
$1/m_Q$ and $1/N_c$ expansion.  The numerical values of the mass combinations 
then can be compared
with the theoretical hierarchy predicted by $1/m_Q$, $1/N_c$ and
$SU(3)$ flavor-breaking suppressions.

Using the derived masses and the measured masses of all the other charmed baryons, it is possible to evaluate the mass hierarchy 
of Table~V.  These numbers are listed in the column labeled Expt in Table~V.  
Good agreement with the theoretical hierarchy is found.
Only the predictions for the $J_\ell^2$ and $J_\ell^i G^{i8}$ mass combinations
and the two mass combinations involving $\Xi_c \Xi_c^\prime$
can be regarded as pure predictions since the choice of parameters 
$m_c$, $\Lambda$,
$(\epsilon \Lambda_\chi)$ and $(\epsilon^2 \Lambda_\chi)$ affects four of
the mass combinations, and
relation~(\ref{mreli}) was imposed to extract the unknown and inaccurately
measured charm baryon masses.  

It also
is worthwhile to study the implications of the predictions Eq.~(\ref{sxo})
for the charm baryon mass spectrum.  The chromomagnetic mass splittings are
evaluated to be
\begin{eqnarray}
&&\left( \Sigma_c^* - \Sigma_c \right) = 79.6 \pm 5.3~{\rm MeV}, \nonumber\\
&&\left( \Xi_c^* - \Xi_c^\prime \right) = 61.1 \pm 3.0~{\rm MeV}, \\
&&\left( \Omega_c^* - \Omega_c \right) = 42.6 \pm 7.3~{\rm MeV}, \nonumber
\end{eqnarray}
which implies that the linear in strangeness 
number contribution to the chromomagnetic
splittings is $\it negative$, so that $\left(\Sigma_c^* - \Sigma_c \right) >
\left(\Xi_c^* - \Xi_c^\prime \right) > \left( \Omega_c^* - \Omega_c \right)$.
This ordering seems reasonable based on intuition from the quark model:
in the quark model, the
hyperfine operator $J_q \cdot J_Q$ is suppressed by $1/m_q m_Q$, which is a
greater suppression for $q=s$ than for $q=u,d$. 
The differences of these splittings are equal as dictated by
Eq.~(\ref{mreli}), and given by $18.5 \pm 4.6$~MeV, so Eq.~(\ref{esr})
which holds at this level is not very accurate.  The spin-averaged
sextet masses are evaluated to be 
\begin{eqnarray}
&&\frac 1 3 \left( \Sigma_c + 2 \Sigma_c^* \right) = 2506.0 \pm 3.5~{\rm MeV},
\nonumber\\
&&\frac 1 3 \left( \Xi_c^\prime + 2 \Xi_c^* \right) = 2623.6 \pm 1.4~{\rm MeV},
\\
&&\frac 1 3 \left( \Omega_c + 2 \Omega_c^* \right) = 2732.4 \pm 4.3~{\rm MeV}.
\nonumber
\end{eqnarray}
The sextet mass differences
\begin{eqnarray}\label{sesr}
&&\frac 1 3 \left( \Xi_c^\prime + 2 \Xi_c^* \right) - \frac 1 3 \left( \Sigma_c
+ 2 \Sigma_c^* \right) = 117.6 \pm 3.8~{\rm MeV}, \nonumber\\
&&\frac 1 3 \left( \Omega_c + 2 \Omega_c^* \right) - \frac 1 3 \left(
\Xi_c^\prime + 2 \Xi_c^* \right) = 108.8 \pm 4.5~{\rm MeV}, 
\end{eqnarray}      
are equal at the $8.8$~MeV level.  The $J_\ell^2$
hyperfine splittings in each strangeness sector are
\begin{eqnarray}
&&\frac 1 3 \left( \Sigma_c + 2 \Sigma_c^* \right) - \Lambda_c = 221.0 \pm
3.6~{\rm MeV}, \nonumber\\
&&\frac 1 3 \left( \Xi_c^\prime + 2 \Xi_c^* \right)- \Xi_c = 155.9 \pm 1.8~{\rm MeV}.
\end{eqnarray}
The difference of these splittings is predicted to be large,
\begin{eqnarray}
&&\left[\frac 1 3 \left( \Sigma_c + 2 \Sigma_c^* \right) - \Lambda_c\right]
-\left[\frac 1 3 \left( \Xi_c^\prime + 2 \Xi_c^* \right)- \Xi_c\right]
\nonumber\\
&&\qquad\qquad= 65.1 \pm 4.0~{\rm MeV} .
\end{eqnarray}

Finally, now that all the charm baryon masses have been determined, it is
possible to return to the problem of predicting the bottom baryon masses from
the charm baryon masses.  The heavy quark spin-violating splittings of the
bottom baryons can be obtained by rescaling the charm baryon mass splittings
by a factor $(m_c/m_b) \sim 0.3$.   
Thus, the chromomagnetic mass splittings of the bottom
baryons are predicted to be
\begin{eqnarray}\label{bchrom}
&&\left( \Sigma_b^* - \Sigma_b \right) \sim 23.8 \pm 1.6~{\rm MeV}, \nonumber\\
&&\left( \Xi_b^* - \Xi_b^\prime \right) \sim 18.3 \pm 0.9~{\rm MeV}, \\
&&\left( \Omega_b^* - \Omega_b \right) \sim 12.8 \pm 2.2~{\rm MeV}, \nonumber
\end{eqnarray}
where
the precise numerical values in Eq.~(\ref{bchrom}) will change outside errors
if a different scale factor is used since the error on the scale factor has not
been taken into account.  For example, using the more
rigorous scale factor $(Z_b/Z_c)(m_c/m_b) \sim 0.24$ reduces the central values
by about $5$~MeV. 
The mass relation Eq.~(\ref{nqmreli}) and the second flavor octet mass relation
of Table VI yield precise
predictions for two additional bottom baryon mass combinations
\begin{eqnarray}\label{bi}
&&\frac 1 6 \left[ \left( \Sigma_b + 2 \Sigma_b^* \right) - 2
\left(\Xi_b^\prime + 2 \Xi_b^* \right) + \left( \Omega_b + 2 \Omega_b^* \right)
\right]\nonumber\\
&&\qquad\qquad= -4.43 \pm 1.5~{\rm MeV}, \nonumber\\
&&\left\{-\frac 5 8
\left( \Lambda_b - \Xi_b \right) \right.\\
&&\left.+ \frac 1 {24} \left[ 3 \left(\Sigma_b + 2
\Sigma_b^*\right) - \left(\Xi^\prime_b + 2 \Xi_b^* \right)
-2 \left(\Omega_b + 2 \Omega_b^*\right)\right]\right\} \nonumber\\
&&\qquad\qquad= 42.89 \pm 2.1~{\rm MeV}, \nonumber  
\end{eqnarray} 
where the errors represent the combined theoretical and experimental
accuracy of each
relation.
There are two additional mass relations which are less accurate, namely
the second and third mass relations in Table~VI, with
theoretical accuracies of about $4.8$ and $5.1$~MeV, respectively:
\begin{eqnarray}\label{bii}
&&\left(\Lambda_b - \Xi_b \right) = -182.7 \pm 4.9~{\rm MeV}, \nonumber\\
&&-\frac 1 3 \left( \Lambda_b + 2 \Xi_b \right) 
+ \frac 1 {18} \left[ 3 \left(
\Sigma_b + 2 \Sigma_b^* \right) + 2 \left( \Xi_b^\prime + 2 \Xi_b^* \right) +
\left( \Omega_b + 2 \Omega_b^* \right) \right] \nonumber\\
&&\qquad\qquad= 176.1 \pm 5.4~{\rm MeV}.
\end{eqnarray}
Combined with the measured $\Lambda_b$ mass, these seven constraints determine
the seven masses $\Xi_b$, $\Sigma_b$, $\Xi_b^\prime$, $\Omega_b$, $\Sigma_b^*$,
$\Xi_b^*$, and $\Omega_b^*$.  The three chromomagnetic splittings were given
already in Eq.~(\ref{bchrom}).  The four mass relations in Eq.~(\ref{bi})
and~(\ref{bii}) together with the $\Lambda_b$ measurement determine the four
spin-averaged mass combinations  
\begin{eqnarray}
&&\frac 1 3 \left( \Lambda_b + 2 \Xi_b \right) = 5744.8 \pm 5.8~{\rm MeV},
\nonumber\\
&&\frac 1 3 \left( \Sigma_b + 2 \Sigma_b^* \right) = 5844.0 \pm 8.9~{\rm
MeV}, \nonumber\\
&&\frac 1 3 \left( \Xi_b^\prime + 2 \Xi_b \right) = 5961.6 \pm 10.8~{\rm MeV},
\\
&&\frac 1 3 \left( \Omega_b + 2 \Omega_b^* \right) = 6070.3 \pm 23.6~{\rm MeV},
\nonumber
\end{eqnarray}
where the precision of the
extraction is limited by the theoretical accuracy of the least accurate mass
relations.  Once two additional bottom baryon masses are measured, a more
precise extraction of the remaining unmeasured masses will be possible using
only the most accurate mass relations in Eq.~(\ref{bi}).  Some linear
combinations of the spin-averaged mass combinations are determined accurately,
however.
For example, the two mass relations in Eq.~(\ref{bi}) imply
\begin{eqnarray}
&&\left[\frac 1 3 \left( \Sigma_b + 2 \Sigma_b^* \right) - \Lambda_b\right]
-\left[\frac 1 3 \left( \Xi_b^\prime + 2 \Xi_b^* \right)- \Xi_b\right]
\nonumber\\
&&\qquad\qquad= 65.1 \pm 3.6~{\rm MeV} .
\end{eqnarray}

\section{Isospin-Violating Mass Splittings}

Isospin-violating mass splittings are analyzed in this section for 
completeness.  Almost all isospin splittings will be at the sub-MeV level,
so isospin symmetry is a very good symmetry for the heavy baryon masses.
Isospin breaking arises due to differences in the $u$ and $d$ quark masses
and electromagnetic charges.  These two sources of isospin breaking will
be denoted by the parameters $\epsilon^\prime$ and $\epsilon^{\prime\prime}$.
Isospin breaking due to quark mass differences is purely $I=1$, whereas the electromagnetic mass splittings are second order in the quark charge
matrix and can be $I=1,2$.  Both sources of isospin breaking produce
comparable mass splittings in QCD, so $\epsilon^\prime \sim
\epsilon^{\prime\prime}$.  All $I=2$ mass splittings are electromagnetic mass
splittings suppressed by $\alpha_{\rm EM}/4\pi$, and will be proportional to
the parameter $(\epsilon^{\prime \prime} \Lambda_\chi)$.  The $I=1$ mass
splittings can arise from either source of isospin breaking.  Since these
effects are comparable, the splittings will be written in terms of the
parameter $(\epsilon^\prime \Lambda_\chi)$.

The $I=1$ flavor octet, $(10 + \overline{10})$ and $27$ mass expansions are
given by        
\begin{eqnarray}\label{sot}
&&M^{(8)} = T^3 + {1 \over N_c} J_\ell^i G^{i3} + {1 \over N_c}
{2 \over { m_Q}} J_Q^i G^{i3} 
+ {1 \over N_c^2}{2 \over { m_Q}} \left( J_\ell \cdot J_Q
\right) T^3, \nonumber\\
&&M^{(10 + \overline{10})} = {1 \over N_c^2}{1 \over m_Q}\left(\{T^3, G^{i8} \}
- \{T^8, G^{i3} \} \right), \\ 
&&M^{(27)} = {1 \over N_c} \{ T^3, T^8 \} 
+ {1 \over N_c^2}{1 \over { m_Q}}J_Q^i \left(\{ T^3, G^{i8} \} 
+ \{ T^8, G^{i3} \}\right), \nonumber
\end{eqnarray}
whereas the $I=2$ flavor-$27$ mass expansion is given by
\begin{eqnarray}\label{sot}
&&M^{(27)} = {1 \over N_c} \{ T^3, T^3 \} 
+ {1 \over N_c^2}{2 \over { m_Q}}J_Q^i \{ T^3, G^{i3} \}.
\end{eqnarray}
The mass combinations corresponding to these operators are tabulated in
Table~VIII for $Q=c$ and Table~IX for $Q=b$.  Table~X combines the operator
matrix element, $1/m_Q$, $1/N_c$ and flavor-breaking suppression factors to 
predict a hierarchy of mass relations.  The numerical values of the theoretical
hierarchy are evaluated for canonical values of the parameters in the columns
labeled $Q=c$ and $Q=b$.  The largest isospin mass splitting is $(\Xi_c^+ -
\Xi_c^0)$ [or $(\Xi_b^0 - \Xi_b^-)$], which are equal and of order a few MeV in
magnitude.  These splittings can be used to determine the isospin breaking
parameter $(\epsilon^\prime \Lambda_\chi)$ which sets the scale of the mass
hierarchy.  Experimentally, $(\Xi_c^0 - \Xi_c^+) =5.2 \pm 2.2$~MeV \cite{pdg}.
The next largest mass splitting is the $I=2$ mass combination
\begin{equation}
\frac 1 {10} \left(\Sigma_c^{++} - 2 \Sigma_c^+ + \Sigma_c^0 \right) +
\frac 2 5 \left( \Sigma_c^{*++} - 2 \Sigma_c^{*+} + \Sigma_c^{*0} \right)
\end{equation}
or the analogous splitting for $Q=b$, which will be about
$3.5 \pm 1.5$~MeV.  The $I=1$ mass splitting
\begin{eqnarray}
&&- \frac 5 8 \left( \Xi_c^+ - \Xi_c^0 \right) + \frac 1 {24} \left[ 2
(\Sigma_c^{++} - \Sigma_c^0 ) + (\Xi_c^{\prime+} - \Xi_c^{\prime 0}) \right]
\nonumber\\
&&+ \frac 1 {12} \left[ 2 (\Sigma_c^{*++} - \Sigma_c^{*0}) + (\Xi_c^{*+} -
\Xi_c^{*0} ) \right] 
\end{eqnarray}    
or the $Q=b$ analogue will both be about $1.0 \pm 0.5$~MeV.  All remaining
isospin mass combinations are predicted to be sub-MeV.  The mass relations
obtained by neglect of these operators are given in Tables~VIII and~IX.  

\section{Conclusions}

It has been shown that there are   
light quark and heavy quark spin-flavor symmetries for baryons containing a
single heavy quark in the combined heavy quark and large $N_c$ limits.
A (spin$\otimes$flavor) operator expansion in $1/N_c$ and $1/m_Q$ has been constructed
for heavy quark baryons at finite $m_Q$ and $N_c$.
Heavy quark spin symmetry is present for $N_c \rightarrow
\infty$ or for $m_Q \rightarrow \infty$, so any violation of heavy quark spin
symmetry is suppressed by $(1/N_c m_Q)$.   
In the presence of heavy quark
flavor symmetry, heavy quark spin-flavor symmetry is present for $N_c
\rightarrow \infty$.  For the physical situation of two heavy flavors
$Q=c$ and $Q=b$, heavy quark flavor symmetry is justified by the heavy quark
limit \cite{iw}.  Heavy quark spin-flavor symmetry $SU(4)_h$ is a better
symmetry for baryons than for mesons because violation of the symmetry is
suppressed by $(1/N_c m_Q)$, rather than $1/m_Q$.  In addition to heavy quark
spin-flavor symmetry, there is a light quark $SU(6)$ spin-flavor
symmetry for heavy quark baryons in the large $N_c$ limit \cite{j}. 

The masses of baryons containing a single heavy quark have been analyzed
in a combined expansion in $1/m_Q$, $1/N_c$ and $SU(3)$ flavor symmetry
breaking.  The naive $1/N_c$ scalings
\begin{eqnarray}
&&\bar\Lambda^{\rm baryon} \sim N_c \bar \Lambda^{\rm meson}, \nonumber\\
&&\lambda_1^{\rm baryon} \sim \lambda_1^{\rm meson}, \nonumber\\
&&\lambda_2^{\rm baryon} \sim {1 \over N_c} \lambda_2^{\rm meson}, 
\end{eqnarray}
work beautifully for the baryon masses.  A mass hierarchy is predicted by the
$1/m_Q$, $1/N_c$ and flavor-breaking expansion.  The most suppressed operators
yield mass relations which are well-satisfied experimentally.  The precision of
the mass relations and the magnitude of mass combinations are predicted by the
expansion.  Heavy baryon
mass splittings have been related to octet and decuplet mass splittings.  The
most accurate mass relations, Eq.~(\ref{mreli}),~(\ref{nqmreli}),
and~(\ref{iii}), have been used
to predict the unmeasured or poorly measured charm baryon masses $\Sigma_c^*$,
$\Xi_c^\prime$ and $\Omega_c^*$ to a precision of several MeV.  A number of
interesting features of the charm baryon mass spectrum are found.  These include: 
(i) the mass splitting
\begin{equation}
\left[\frac 1 3 \left(\Sigma_c + 2 \Sigma_c^* \right) - \Lambda_c \right]
-\left[\frac 1 3 \left(\Xi_c^\prime + 2 \Xi_c^* \right) - \Xi_c \right]
\end{equation}
is predicted to be large, (ii) the chromomagnetic hyperfine mass splittings satisfy
a reverse hierarchy,
\begin{equation}
(\Sigma_c^* - \Sigma_c) > (\Xi_c^* - \Xi_c^\prime) > (\Omega_c^* - \Omega_c) ,
\end{equation}
and (iii) the equal spacing rule \cite{savage}
\begin{equation}
(\Sigma_c^* - \Sigma_c) - (\Xi_c^* - \Xi_c^\prime) =
(\Xi_c^* - \Xi_c^\prime) - (\Omega_c^* - \Omega_c)
\end{equation}
is almost exact.       
Bottom baryon mass splittings are predicted in terms of charm baryon mass
splittings and mass splittings of the octet and decuplet baryons.  Some bottom
baryon mass splittings are predicted very accurately.

\vskip 1in
As this work neared completion, manuscript\cite{falk} appeared.
The present analysis shows that Eq.~(2d) of \cite{falk}
is not well satisfied.

Noted added in proof:  There are no renormalon ambiguities in the operator
coefficients which correspond to physical mass splittings.  The renormalon
ambiguity of the heavy quark mass $m_Q$ and $\bar \Lambda^{\rm baryon} \sim
N_c \Lambda_{\rm QCD}$ is of order $\Lambda_{\rm QCD}$, and formally appears
as a term $N_Q \Lambda_{\rm QCD}$ in the expansion (4.13).

The two very accurate mass relations Eqs. (5.7) and (5.15)
yield very precise predictions for $\Xi_c^\prime$ and 
$(\Sigma_c^* + \Omega_c^*)$.  A new CLEO measurement of 
$\Sigma_c^* = 2518.6 \pm 2.2$ MeV is significantly lower than
previous measurements.  In addition, there is a somewhat more
accurate value for $\Omega_c = 2699.9 \pm 2.9$ MeV from the E687
Collaboration.  The more accurate $\Omega_c$ value changes the 
$\Xi_c^\prime$ prediction slightly to
$$ \Xi_c^\prime = 2580.8 \pm 2.1 {\rm MeV}.  $$
The lower $\Sigma_c^*$ mass implies a larger value for the predicted
$\Omega_c^*$ mass,
$$ \Omega_c^* = 2760.6 \pm 6.4 {\rm MeV}.   $$
The mass combination quoted in Eqs. (5.20)--(5.29) will be modified
accordingly.  The only significant modification is that the
differences of the chromomagnetic mass splittings 
$(\Sigma_Q^* - \Sigma_Q)$, $(\Xi_Q^* - \Xi_Q^\prime)$ and 
$(\Omega_Q^* - \Omega_Q)$ are much smaller than before, which
implies that relation (5.13) is as small as possible.  This occurs if
the magnitude of the first mass combination in Eq. (5.12) is 
dominated by the $\Xi_Q \Xi_Q^\prime$ mixing term.

\section*{Acknowledgments}

I wish to thank M. Luke for a valuable discussion, and
R. Kutschke, and V. Sharma for information on the 
experimental data.  

This work was supported in part by the Department of Energy
under grant DOE-FG03-90ER40546.  E.J. also was supported in part by NYI
award PHY-9457911 from the National Science Foundation and by a research
fellowship from the Alfred P. Sloan Foundation.

\vfill\break\eject

\onecolumn 

\widetext

\centerline{\bf Figure Captions}

\begin{figure}
\caption{Baryon representation of $SU(2 N_F) \times SU(2)_Q \times
SU(2)_{Q^\prime}$ spin-flavor symmetry for large finite $N_c$.  The total
number of boxes equals $N_c = N_\ell + N_h$ where $N_h = N_Q + N_{Q^\prime}$.}
\end{figure}

\def\nbox{\hbox{$\bsqr\bsqr\raise2.7pt\hbox{$\,\cdot\cdot
\cdot\cdot\cdot\,$}\bsqr$}}

\centerline{$\underbrace{\nbox}_{N_\ell}$ $\otimes$ 
$\underbrace{\nbox}_{N_Q}$ 
$\otimes$ $\underbrace{\nbox}_{N_{Q^\prime}}$}
\bigskip
\centerline{Figure 1}
\vskip1in

\vfill\break\eject

\begin{table}[htbp]
\caption{Mass splittings of baryons containing a single heavy quark $Q$ for
strangeness $S=0$ and $S=-1$ baryons.  Each operator is an isospin singlet; the
$J_\ell \cdot J_Q$ operator violates heavy quark spin symmetry.  The operator
matrix element and $1/m_Q$, $1/N_c$ and isospin flavor-breaking suppressions of
each mass combination are tabulated.  
The singlet mass combination has a
contribution $m_Q$ and a contribution of order $N_c$ at leading order,
i.e. $m_Q + N_c \Lambda$.} 
\smallskip
\label{tab:one}
\centerline{\vbox{ \tabskip=0pt \offinterlineskip
\def\tablerule{\noalign{\hrule}}
\def\space{height 2pt&\omit&&\omit&&\omit&&\omit&&\omit&&\omit&&
\omit&&\omit\cr}
\hrule \vskip 2pt
\halign{
\vrule #&\strut\hfil\ #\hfil&&
\vrule #&\strut\hfil\ #\ \hfil\cr
\tablerule\space
& Operator && $(SU(2), J_Q)$ && \hfil  Mass Combination \hfil && 
$\langle {\cal O} \rangle$ &&
$1/m_Q$ && $1/N_c$ && Flavor &&\omit\cr
\tablerule\space\space
& $\openone$ && $(1,0)$ && $\Lambda_Q$  
&& $1$ && * && * && $1$ 
&&\omit\cr\space\space
& $J_\ell^2$ && $(1,0)$ && 
$\frac 1 3 \left( \Sigma_Q + 2 \Sigma_Q^* \right)-\Lambda_Q$  
&& $2$ && $1$ && $1/N_c$ && $1$
&&\omit\cr\space\space
& $J_\ell \cdot J_Q$ && $(1,1)$ && $\Sigma^*_Q - \Sigma_Q$  
&& $\frac 3 2$ && $2/ m_Q$ && $1/N_c$ && $1$
&&\omit\cr\space\space
\tablerule\space\space
& $\openone$ && $(1,0)$ && $\Xi_Q$  
&& $1$ && * && * && $1$ 
&&\omit\cr\space\space
& $J_\ell^2$ && $(1,0)$ && 
$\frac 1 3 \left( \Xi_Q^\prime + 2 \Xi_Q^* \right)-\Xi_Q$  
&& $2$ && $1$ && $1/N_c$ && $1$
&&\omit\cr\space\space
& $J_\ell \cdot J_Q$ && $(1,1)$ && $\Xi^*_Q - \Xi^\prime_Q$  
&& $\frac 3 2$ && $2/m_Q$ && $1/N_c$ && $1$
&&\omit\cr\space\space
\space\tablerule
}
\vskip 2pt \hrule
}}
\end{table}

\begin{table}[htbp]
\caption{$I_h = 1$ heavy baryon mass splittings which violate 
heavy quark flavor symmetry, but preserve heavy quark spin symmetry
for strangeness $S=0$ and $S=-1$ baryons.  
The mass splittings are suppressed by $\left( 1/2 m_c - 1/2 m_b \right)$
and one additional factor of $1/N_c$.  The $I_h^3$ mass combination has
a leading contribution of $\left(m_c - m_b\right)$.}
\smallskip
\label{tab:three}
\centerline{\vbox{ \tabskip=0pt \offinterlineskip
\def\tablerule{\noalign{\hrule}}
\def\space{height 2pt&\omit&&\omit&&\omit&&\omit&&\omit&&\omit&&
\omit&&\omit\cr}
\hrule \vskip 2pt
\halign{
\vrule #&\strut\hfil\ #\hfil&&
\vrule #&\strut\hfil\ #\ \hfil\cr
\tablerule\space
& Operator && $(SU(2), J_Q)$ && \hfil Mass Combination \hfil && 
$\langle {\cal O} \rangle$ && $1/m_Q$ &&
$1/N_c$ && Flavor &&\omit\cr
\tablerule\space\space\space\space
& $I_h^3$ && $(1,0)$ && $\left(\Lambda_c -\Lambda_b \right)$  
&& $1$ && $\left( m_c - m_b \right)$ && $1$ && $1$ 
&&\omit\cr\space\space\space\space
& $I_h^3 \left( J_\ell\right)^2$ && $(1,0)$ 
&& $\left[ \frac 1 3 \left( \Sigma_c + 2 \Sigma_c^* \right)-\Lambda_c\right]
-\left[ \frac 1 3 \left( \Sigma_b + 2 \Sigma_b^* \right)-\Lambda_b\right]$  
&& $2$ && $\left({1 \over {2m_c}} - {1 \over {2m_b}}\right)$ && ${1 / N_c^2}$ && $1$
&&\omit\cr\space\space\space
\tablerule\space\space\space\space
& $I_h^3$ && $(1,0)$ && $\left(\Xi_c -\Xi_b \right)$  
&& $1$ && $\left( m_c - m_b \right)$ && $1$ && $1$ 
&&\omit\cr\space\space\space\space
& $I_h^3 \left( J_\ell\right)^2$ && $(1,0)$ 
&& $\left[ \frac 1 3 \left( \Xi_c^\prime + 2 \Xi_c^* \right)-\Xi_c\right]
-\left[ \frac 1 3 \left( \Xi^\prime_b + 2 \Xi_b^* \right)-\Xi_b\right]$  
&& $2$ && $\left({1 \over {2m_c}} - {1 \over {2m_b}}\right)$ && ${1 / N_c^2}$
&& $1$ 
&&\omit\cr\space\space\space
\space\tablerule
}
\vskip 2pt \hrule
}}
\end{table}

\begin{table}[htbp]
\caption{Mass splittings between baryons containing a single heavy quark $Q$
and baryons containing no heavy quarks for strangeness 
$S=0$ and $S=-1$ baryons.  The mass splittings are suppressed by an additional
factor of $1/ N_c$.
The $N_Q$ mass combination has a leading order contribution of $m_Q$.}
\smallskip
\label{tab:one}
\centerline{\vbox{ \tabskip=0pt \offinterlineskip
\def\tablerule{\noalign{\hrule}}
\def\space{height 2pt&\omit&&\omit&&\omit&&\omit&&\omit&&\omit&&
\omit&&\omit\cr}
\hrule \vskip 2pt
\halign{
\vrule #&\strut\hfil\ #\hfil&&
\vrule #&\strut\hfil\ #\ \hfil\cr
\tablerule\space
& Operator && $(SU(2), J_Q)$ && \hfil  Mass Combination \hfil && 
$\langle {\cal O} \rangle$ &&
$1/m_Q$ && $1/N_c$ && Flavor &&\omit\cr
\tablerule\space\space\space\space
& $N_Q$ && $(1,0)$ && $\Lambda_Q 
- \frac 1 4 \left( 5 N - \Delta \right)$  
&& $1$ && $m_Q$ && 1 && $1$ 
&&\omit\cr\space\space\space\space
& $N_Q J_\ell^2$ && $(1,0)$ && $\left[ \frac 1 3 \left( \Sigma_Q + 2 \Sigma_Q^*
\right) - \Lambda_Q \right] - \frac 2 3 \left( \Delta - N \right)$  
&& $2$ && $1$ && $1/N_c^2$ && $1$
&&\omit\cr\space\space\space\space
\tablerule\space\space\space\space
& $N_Q$ && $(1,0)$ && $\Xi_Q 
- \frac 1 4 \left[ \frac 5 4 \left( 3 \Sigma + \Lambda \right) 
- \Sigma^*\right]$  
&& $1$ && $m_Q$ && 1 && $1$ 
&&\omit\cr\space\space\space\space
& $N_Q J_\ell^2$ && $(1,0)$ && $\left[ \frac 1 3 \left( \Xi^\prime_Q 
+ 2 \Xi_Q^* \right) - \Xi_Q \right] 
- \frac 2 3 \left[ \Sigma^* 
- \frac 1 4 \left( 3\Sigma + \Lambda \right)\right]$  
&& $2$ && $1$ && $1/N_c^2$ && $1$
&&\omit\cr\space\space\space\space
\space\tablerule
}
\vskip 2pt \hrule
}}
\end{table}

\vfill\break\eject

\vfill\break\eject

\begin{table}[htbp]
\caption{Mass splittings of baryons containing a single heavy quark $Q$.
$SU(3)$ flavor and heavy quark spin $J_Q$ symmetry quantum numbers of each
operator and mass combination are given explicitly.  The operator matrix
element for each mass combination and suppression factors of $1/m_Q$, $1/N_c$
and $SU(3)$ flavor-breaking $\epsilon$ are tabulated.  The singlet operator is
$m_Q + N_c \Lambda$ at leading order.}
\smallskip
\label{tab:one}
\centerline{\vbox{ \tabskip=0pt \offinterlineskip
\def\tablerule{\noalign{\hrule}}
\def\space{height 2pt&\omit&&\omit&&\omit&&\omit&&\omit&&\omit&&
\omit&&\omit\cr}
\hrule \vskip 2pt
\halign{
\vrule #&\strut\hfil\ #\hfil&&
\vrule #&\strut\hfil\ #\ \hfil\cr
\tablerule\space
& Operator && $(SU(3), J_Q)$ && \hfil  Mass Combination \hfil && 
$\langle {\cal O} \rangle$ &&
$1/m_Q$ && $1/N_c$ && Flavor &&\omit\cr
\tablerule\space\space
& $\openone$ && $(1,0)$ && $\frac 1 3 \left(\Lambda_Q + 2 \Xi_Q \right)$  
&& $1$ && * && * && $1$ 
&&\omit\cr\space\space
& $J_\ell^2$ && $(1,0)$ && 
$-\frac 1 3 \left(\Lambda_Q + 2 \Xi_Q \right)
+ \frac 1 {18} \left[ 3 \left( \Sigma_Q + 2 \Sigma_Q^* \right)
+ 2 \left( \Xi^\prime_Q + 2 \Xi_Q^* \right) 
+ \left( \Omega_Q  + 2 \Omega_Q^* \right) \right]$  
&& $2$ && $1$ && $1/N_c$ && $1$
&&\omit\cr\space\space
& $J_\ell \cdot J_Q$ && $(1,1)$ && 
$\frac 1 6 \left[ 3 \left( \Sigma^*_Q - \Sigma_Q \right)
+ 2 \left( \Xi^*_Q - \Xi_Q^\prime \right) 
+ \left( \Omega^*_Q - \Omega_Q \right) \right]$   
&& $\frac 3 2$ && $2/m_Q$ && $1/N_c$ && $1$
&&\omit\cr\space\space
& $T^8$ && $(8,0)$ && $\left( \Lambda_Q - \Xi_Q \right)$  
&& $\frac 3 {2 \sqrt{3}}$ && $1$ && $1$ && $\epsilon$
&&\omit\cr\space\space
& $J_\ell^i G^{i8}$ && $(8,0)$ && 
$-\frac 5 8
\left( \Lambda_Q - \Xi_Q \right)+ \frac 1 {24} \left[ 3 \left( \Sigma_Q
+ 2 \Sigma_Q^* \right)  - \left( \Xi^\prime_Q + 2 \Xi_Q^* \right)
-2 \left( \Omega_Q + 2 \Omega_Q^* \right) \right] $    
&& $\frac 1 {2 \sqrt{3}} \frac {15} 8$ && $1$ && $1/N_c$ && $\epsilon$
&&\omit\cr\space\space
& $J_Q^i G^{i8}$ && $(8,1)$ && $\Xi_Q \Xi^\prime_Q$  
&& $\frac 3 8$ && $2/m_Q$ && $1/N_c$ && $\epsilon$
&&\omit\cr\space\space
& $\left( J_\ell \cdot J_Q \right) T^8$ && $(8,1)$ && 
$\frac 1 {6} \left[ 3 \left( \Sigma^*_Q - \Sigma_Q \right) 
- \left( \Xi^*_Q - \Xi^\prime_Q \right)
-2 \left( \Omega^*_Q - \Omega_Q \right) \right] 
-\frac 5 {2\sqrt{3}} \Xi_Q \Xi^\prime_Q $   
&& $\frac 1 {2 \sqrt{3}} \frac {15} 4$ && $2/m_Q$ && $1/N_c^2$ && $\epsilon$
&&\omit\cr\space\space
& $\{T^8,T^8\}$ && $(27,0)$ && 
$\frac 1 {6} \left[ \left( \Sigma_Q + 2 \Sigma_Q^* \right)
- 2 \left( \Xi^\prime_Q + 2 \Xi_Q^* \right)
+ \left( \Omega_Q + 2 \Omega_Q^* \right) \right] $  
&& $\frac 3 2$ && $1$ && $1/N_c$ && $\epsilon^2$
&&\omit\cr\space\space
& $J_Q^i \{T^8,G^{i8}\}$ && $(27,1)$ && $\frac 1 4 \left[ \left( \Sigma^*_Q - \Sigma_Q \right)
- 2 \left( \Xi^*_Q - \Xi_Q^\prime \right) 
+ \left( \Omega^*_Q - \Omega_Q \right) \right]$   
&& $\frac 9 {16}$ && $2/m_Q$ && $1/N_c^2$ && $\epsilon^2$
&&\omit\cr\space\space
\space\tablerule
}
\vskip 2pt \hrule
}}
\end{table}

\begin{table}[htbp]
\caption{Heavy baryon mass hierarchy for $Q=c$ and $Q=b$.  The experimental
charm baryon mass splittings are evaluated using theoretically
extracted values for the 
three masses $\Sigma_c^* = 2532.5 \pm 5.3$~MeV, $\Xi_c^\prime = 2582.9 \pm
2.5$~MeV and $\Omega_c^* = 2746.6 \pm 6.1$~MeV.  Agreement between the theory
values appearing with an asterisk and experiment is imposed by the
determination of the parameter values $m_c$, $\Lambda$, $(\epsilon \Lambda_\chi
)$ and $(\epsilon^2 \Lambda_\chi)$.  Comparison of the remaining theory and
experimental values provides support for the theoretical hierarchy.
}
\smallskip
\label{tab:two}
\centerline{\vbox{ \tabskip=0pt \offinterlineskip
\def\tablerule{\noalign{\hrule}}
\def\space{height 2pt&\omit&&\omit&&\omit&&\omit&&\omit&&\omit\cr}
\hrule \vskip 2pt
\halign{
\vrule #&\strut\hfil\ #\hfil&&
\vrule #&\strut\hfil\ #\ \hfil\cr
\tablerule\space
& \hfil  Mass Combination \hfil && Theory && $Q=c$ && Expt $Q=c$ 
&& $Q=b$ &&\omit\cr
\tablerule\space\space
& $\frac 1 3 \left(\Lambda_Q + 2 \Xi_Q \right)$  
&& $m_Q + N_c \Lambda$ && $2380^*$  && $2406.8 \pm 0.8$ && $5687$  
&&\omit\cr\space\space
& $-\frac 1 3 \left(\Lambda_Q + 2 \Xi_Q \right)
+ \frac 1 {18} \left[ 3 \left( \Sigma_Q + 2 \Sigma_Q^* \right)
+ 2 \left( \Xi^\prime_Q + 2 \Xi_Q^* \right) 
+ \left( \Omega_Q  + 2 \Omega_Q^* \right) \right]$  
&& $2 {1 \over N_c} \Lambda$ && $207$ && $176.1 \pm 1.9$ && $207$ 
&&\omit\cr\space\space
& $\frac 1 6 \left[ 3 \left( \Sigma^*_Q - \Sigma_Q \right)
+ 2 \left( \Xi^*_Q - \Xi_Q^\prime \right) 
+ \left( \Omega^*_Q - \Omega_Q \right) \right]$  
&& ${3}{1 \over N_c}{\Lambda^2 \over m_Q}$ && $66^*$ && $67.3 \pm 3.1$ && $20$
&&\omit\cr\space\space
& $\left( \Lambda_Q - \Xi_Q \right)$  
&& $\frac 3 {2 \sqrt{3}} (\epsilon \Lambda_\chi)$ && $-195^*$ && 
$-182.7 \pm 1.3$ && $-195$
&&\omit\cr\space\space
& $-\frac 5 8
\left( \Lambda_Q - \Xi_Q \right)+ \frac 1 {24} \left[ 3 \left( \Sigma_Q
+ 2 \Sigma_Q^* \right)  - \left( \Xi^\prime_Q + 2 \Xi_Q^* \right)
-2 \left( \Omega_Q + 2 \Omega_Q^* \right) \right] $  
&& $\frac 1 {2\sqrt{3}}\frac {15} 8 {1 \over N_c}(\epsilon \Lambda_\chi)$ &&
$40.6$ && $42.9 \pm 1.9$ && $40.6$
&&\omit\cr\space\space
& $\Xi_Q \Xi^\prime_Q $  
&& $\frac 3 4 {1 \over N_c} 
{\Lambda \over m_Q} (\epsilon \Lambda_\chi)$ && $11.8$ && $--$ && $3.5$
&&\omit\cr\space\space
& $\frac 1 {6} \left[ 3 \left( \Sigma^*_Q - \Sigma_Q \right) 
- \left( \Xi^*_Q - \Xi^\prime_Q \right)
-2 \left( \Omega^*_Q - \Omega_Q \right) \right] 
-\frac 5 {2\sqrt{3}} \Xi_Q \Xi^\prime_Q $  
&& $\frac 1 {2\sqrt{3}} \frac {15} 2 {1 \over N_c^2} 
{\Lambda \over m_Q} (\epsilon \Lambda_\chi)$
&& $11.3$ && $(15.4 \pm 3.6) - \frac 5 {2\sqrt{3}} \Xi_c \Xi^\prime_c $ &&
$3.4$ 
&&\omit\cr\space\space
& $\frac 1 {6} \left[ \left( \Sigma_Q + 2 \Sigma_Q^* \right)
- 2 \left( \Xi^\prime_Q + 2 \Xi_Q^* \right)
+ \left( \Omega_Q + 2 \Omega_Q^* \right) \right] $  
&& $\frac 3 2 {1 \over N_c} (\epsilon^2 \Lambda_\chi)$ && $-4.4^*$ && $-4.4 \pm
3.1$ && $-4.4$
&&\omit\cr\space\space
& $\frac 1 4 \left[ \left( \Sigma^*_Q - \Sigma_Q \right)
- 2 \left( \Xi^*_Q - \Xi_Q^\prime \right) 
+ \left( \Omega^*_Q - \Omega_Q \right) \right]$  
&& $\frac 9 8 {1 \over N_c^2} {\Lambda \over m_Q}
(\epsilon^2 \Lambda_\chi)$ && $0.23$ && $0 \pm 2.7$ && $0.07$
&&\omit\cr\space\space
\space\tablerule
}
\vskip 2pt \hrule
}}
\end{table}

\vfill\break\eject

\begin{table}[htbp]
\caption{$I_h = 1$ heavy baryon mass splittings which violate 
heavy quark flavor symmetry, but preserve heavy quark spin symmetry.  
The mass splittings are suppressed by $\left( 1/ 2m_c - 1/ 2m_b \right)$
and one additional factor of $1/N_c$.  The $I_h^3$ mass combination has
a leading contribution of $\left(m_c - m_b\right)$. }
\smallskip
\label{tab:three}
\centerline{\vbox{ \tabskip=0pt \offinterlineskip
\def\tablerule{\noalign{\hrule}}
\def\space{height 2pt&\omit&&\omit&&\omit&&\omit&&\omit&&\omit&&
\omit&&\omit\cr}
\hrule \vskip 2pt
\halign{
\vrule #&\strut\hfil\ #\hfil&&
\vrule #&\strut\hfil\ #\ \hfil\cr
\tablerule\space
& Operator && $(SU(3), J_Q)$ && \hfil Mass Combination \hfil && 
$\langle {\cal O} \rangle$ && $1/m$ &&
$1/N_c$ && Flavor &&\omit\cr
\tablerule\space\space\space\space
& $I_h^3$ && $(1,0)$ && $\frac 1 3 \left(\Lambda_c + 2 \Xi_c \right)-\frac 1 3 \left(\Lambda_b + 2 \Xi_b \right)$  
&& $1$ && $\left( m_c - m_b \right)$ && $1$ && $1$ 
&&\omit\cr\space\space\space\space
& $I_h^3 \left( J_\ell\right)^2$ && $(1,0)$ 
&& $\left[ -\frac 1 3 \left(\Lambda_c + 2 \Xi_c \right)
+ \frac 1 {18} \left( 3 \Sigma_c + 2 \Xi^\prime_c + \Omega_c \right)
+ \frac 1 {9} \left( 3 \Sigma^*_c + 2 \Xi^*_c + \Omega^*_c \right)\right]$  
&& $2$ && $\left({1 \over {2m_c}} - {1 \over {2m_b}}\right)$ && ${1 / N_c^2}$
&& $1$ 
&&\omit\cr
&  &&  && $ 
\qquad-\left[-\frac 1 3 \left(\Lambda_b + 2 \Xi_b \right)
+ \frac 1 {18} \left( 3 \Sigma_b + 2 \Xi^\prime_b + \Omega_b \right)
+ \frac 1 {9} \left( 3 \Sigma^*_b + 2 \Xi^*_b + \Omega^*_b \right)\right]$  
&& &&  &&  && 
&&\omit\cr\space\space\space\space
& $I_h^3 T^8$ && $(8,0)$ && $
\left( \Lambda_c - \Xi_c \right) 
-  \left( \Lambda_b - \Xi_b \right)$  
&& ${3 \over {2\sqrt{3}}}$
&& $\left({1 \over {2m_c}} - {1 \over {2m_b}}\right)$ && $1/N_c$ && $\epsilon$
&&\omit\cr\space\space\space\space
& $I_h^3 J_\ell^i G^{i8}$ && $(8,0)$ && $\left[ -\frac 5 8
\left( \Lambda_c - \Xi_c \right)+ \frac 1 {24} \left( 3 \Sigma_c - \Xi^\prime_c
-2 \Omega_c \right) + \frac 1 {12} \left(3 \Sigma^*_c - \Xi^*_c -2 \Omega^*_c \right) \right]$  
&& ${1 \over {2\sqrt{3}} } {{15} \over 8}$
&& $\left({1 \over {2m_c}} - {1 \over {2 m_b}}\right)$ && $1/N_c^2$ &&
$\epsilon$ 
&&\omit\cr
& && && $\qquad -\left[ -\frac 5 8
\left( \Lambda_b - \Xi_b \right)+ \frac 1 {24} \left( 3 \Sigma_b - \Xi^\prime_b
-2 \Omega_b \right) + \frac 1 {12} \left(3 \Sigma^*_b - \Xi^*_b -2 \Omega^*_b \right) \right]$  
&& && && &&
&&\omit\cr\space\space\space\space
& $I_h^3 \{T^8,T^8\}$ && $(27,0)$ && $\left[\frac 1 6 \left( \Sigma_c - 2 \Xi^\prime_c +
\Omega_c \right) + \frac 1 3 \left( \Sigma^*_c - 2 \Xi^*_c 
+ \Omega^*_c \right)\right]$  
&& $3 \over 2$
&& $\left({1 \over {2m_c}} - {1 \over {2 m_b}}\right)$ && $1/N_c^2$ &&
$\epsilon^2$ 
&&\omit\cr
& && && $\qquad -\left[ \frac 1 6 \left( \Sigma_b - 2 \Xi^\prime_b +
\Omega_b \right) + \frac 1 3 \left( \Sigma^*_b - 2 \Xi^*_b 
+ \Omega^*_b \right)\right]$  
&& && && &&
&&\omit\cr\space\space\space\space
\space\tablerule
}
\vskip 2pt \hrule
}}
\end{table}

\vfill\break\eject

\begin{table}[htbp]
\caption{Mass splittings between baryons containing a single heavy quark $Q$
and baryons containing no heavy quarks.  The mass splittings are suppressed by
an additional factor of $1/ N_c$.
The $N_Q$ mass combination has a leading order contribution of $m_Q$.}
\smallskip
\label{tab:one}
\centerline{\vbox{ \tabskip=0pt \offinterlineskip
\def\tablerule{\noalign{\hrule}}
\def\space{height 2pt&\omit&&\omit&&\omit&&\omit&&\omit&&\omit&&
\omit&&\omit\cr}
\hrule \vskip 2pt
\halign{
\vrule #&\strut\hfil\ #\hfil&&
\vrule #&\strut\hfil\ #\ \hfil\cr
\tablerule\space
& Operator && $(SU(3), J_Q)$ && \hfil  Mass Combination \hfil && 
$\langle {\cal O} \rangle$ &&
$1/m_Q$ && $1/N_c$ && Flavor &&\omit\cr
\tablerule\space\space\space\space
& $N_Q$ && $(1,0)$ && $\frac 1 3 \left(\Lambda_Q + 2 \Xi_Q \right)
- \frac 1 4 \left[ \frac 5 8 \left( 2 N + 3 \Sigma + \Lambda + 2 \Xi \right) 
-\frac 1 {10} \left( 4 \Delta + 3 \Sigma^* + 2 \Xi^* +\Omega \right) \right]$  
&& $1$ && $m_Q$ && 1 && $1$ 
&&\omit\cr\space\space\space\space
& $N_Q J_\ell^2$ && $(1,0)$ && $\left\{-\frac 1 3 \left(\Lambda_Q 
+ 2 \Xi_Q \right)+ \frac 1 {18} \left[ 3 \left(\Sigma_Q + 2 \Sigma_Q^*\right) 
+ 2 \left( \Xi^\prime_Q + 2 \Xi^*_Q \right) 
+ \left( \Omega_Q + 2 \Omega_Q^*\right)\right]\right\}$  
&& $2$ && $1$ && $1/N_c^2$ && $1$
&&\omit\cr\space
&  &&  && $- \frac 2 3 \left[ \frac 1 {10} \left( 4 \Delta + 3 \Sigma^* + 2
\Xi^* + \Omega \right) - \frac 1 8 \left( 2 N + 3 \Sigma + \Lambda + 2 \Xi
\right) \right]$  
&&  &&  &&  &&
&&\omit\cr\space\space\space\space
& $N_Q T^8$ && $(8,0)$ && $\left( \Lambda_Q - \Xi_Q \right)- \frac 1 {8}
\left( 6 N - 3 \Sigma +  \Lambda - 4 \Xi \right) + \frac 1 {20} \left( 2 \Delta
- \Xi^* - \Omega \right)$  
&& $\frac 3 {2 \sqrt{3}}$ && $1$ && $1/N_c$ && $\epsilon$
&&\omit\cr\space\space\space\space
& $N_Q J_\ell^i G^{i8}$ && $(8,0)$ && $\left\{-\frac 5 8
\left( \Lambda_Q - \Xi_Q \right)+ \frac 1 {24} \left[ 3 \left(\Sigma_Q + 2
\Sigma_Q^*\right) - \left(\Xi^\prime_Q + 2 \Xi_Q^* \right)
-2 \left(\Omega_Q + 2 \Omega_Q^*\right)\right]\right\}$  
&& $\frac 1 {2 \sqrt{3}} \frac {15} 8$ && $1$ && $1/N_c^2$ && $\epsilon$
&&\omit\cr\space
&  && && $+\frac 1 {24} \left( 8 N - 9 \Sigma + 3 \Lambda -2 \Xi \right)
- \frac 1 {12} \left( 2 \Delta - \Xi^* - \Omega \right)$  
&&  &&  &&  && 
&&\omit\cr\space\space\space\space
& $N_Q \{T^8,T^8\}$ && $(27,0)$ && $\frac 1 6 \left[\left( \Sigma_Q + 2
\Sigma_Q^* \right)- 2 \left(\Xi^\prime_Q + 2 \Xi_Q^* \right) +
\left(\Omega_Q + 2 \Omega_Q^* \right) \right]$  
&& $\frac 3 2$ && $1$ && $1/N_c^2$ && $\epsilon^2$
&&\omit\cr\space
&  &&  && $+\frac 1 3 \left[ -\frac 1 4 \left( 2 N - \Sigma -3 \Lambda + 2 \Xi
\right) - \frac 1 7 \left( 4 \Delta - 5 \Sigma ^* -2 \Xi^* + 3 \Omega \right) \right]$  
&&  &&  &&  && 
&&\omit\cr\space\space\space\space
\space\tablerule
}
\vskip 2pt \hrule
}}
\end{table}

\vfill\break\eject

\begin{table}[htbp]
\caption{Isospin-violating mass splittings of baryons containing a single
charm quark.}
\smallskip
\label{tab:four}
\centerline{\vbox{ \tabskip=0pt \offinterlineskip
\def\tablerule{\noalign{\hrule}}
\def\space{height 2pt&\omit&&\omit&&\omit&&\omit
&&\omit&&\omit&&\omit&&\omit\cr}
\hrule \vskip 2pt
\halign{
\vrule #&\strut\hfil\ #\hfil&&
\vrule #&\strut\hfil\ #\ \hfil\cr
\tablerule\space
& Operator && $(SU(3), J_c)$ && \hfil  Mass Combination \hfil && 
$\langle {\cal O} \rangle$ && $1/m_c$ &&
$1/N_c$ && Flavor &&\omit\cr
\tablerule\space\space\space
& \omit && \omit &&\hfil $I=1$ \hfil &&  && && && &&\omit\cr
\space\space\space\space
& $T^3$ && $(8,0)$ && $\left(\Xi_c^+ - \Xi_c^0 \right)$  
&& $1$ && $1$  && $1$  && $\epsilon^\prime$ 
&&\omit\cr\space\space\space\space\space
& $J_\ell^i G^{i3}$ && $(8,0)$ && 
$-\frac 5 8 \left( \Xi^+_c - \Xi^0_c \right) 
+ \frac 1 {24} \left[ 2 \left( \Sigma^{++}_c - \Sigma^0_c \right)
+ \left( \Xi^{\prime +}_c - \Xi^{\prime 0}_c \right) \right]$  
&& $\frac 5 8$ && $1$ && $1/N_c$ && $\epsilon^\prime$
&&\omit\cr\space
&  &&  && $+ \frac 1 {12} \left[ 2 \left( \Sigma^{*++}_c 
- \Sigma^{*0}_c \right)
+ \left( \Xi^{*+}_c - \Xi^{*0}_c \right) \right]$  
&& &&  &&  && 
&&\omit\cr\space\space\space\space\space
& $J_c^i G^{i3}$ && $(8,1)$ && $\frac 1 2  \left[
\Lambda_c^+\Sigma_c^+ + \frac 1 2 \left( \Xi^+_c \Xi^{\prime +}_c - \Xi^0_c
\Xi^{\prime 0}_c \right) \right]$  
&& $\frac {3 \sqrt{3}} 8$ && $2/m_c$ && $1/N_c$ && $\epsilon^\prime$
&&\omit\cr\space\space\space\space\space
& $\left( J_\ell \cdot J_c \right) T^3$ && $(8,1)$ && $-\frac 1 6
\left[ 2 \left(\Sigma_c^{++} - \Sigma_c^0 \right) + \left(\Xi_c^{\prime +}
-\Xi_c^{\prime 0} \right) \right]$
&& $\frac 5 4$ && $2/m_c$ && $1/N_c^2$ && $\epsilon^\prime$
&&\omit\cr\space
&  &&  && $\qquad-{5 \over {3\sqrt{3}} } \left[
\Lambda_c^+\Sigma_c^+ + \frac 1 2 \left( \Xi^+_c \Xi^{\prime +}_c - \Xi^0_c
\Xi^{\prime 0}_c \right) \right]$ 
&& &&  &&  &&
&&\omit\cr\space
&  &&  && $\qquad\qquad + \frac 1 {6} \left[ 
2 \left(\Sigma_c^{*++} - \Sigma_c^{*0} \right) 
+ \left(\Xi_c^{* +} -\Xi_c^{* 0} \right) \right] $  
&& &&  &&  && 
&&\omit\cr\space\space\space\space\space
& $\{T^3,T^8\}$ && $(27,0)$ && $
\frac 1 {18} \left[ \left( \Sigma_c^{++} - \Sigma_c^0 \right) 
- 2 \left( \Xi^{\prime +}_c -\Xi^{\prime 0}_c \right) \right]$
&& $\frac 1 {\sqrt{3}}$ && $1$ && $1/N_c$ && $\epsilon\epsilon^\prime$
&&\omit\cr\space
&  &&  && $\qquad+
\frac 1 9 \left[ \left( \Sigma_c^{*++} - \Sigma_c^{*0} \right) 
- 2 \left( \Xi^{* +}_c -\Xi^{* 0}_c \right) \right]$
&& && && && 
&&\omit\cr\space\space\space\space\space
& $J_c^i \left( \{T^3,G^{i8} \} + \{T^8,G^{i3} \} \right)$ 
&& $(27,1)$ && $-
\frac 1 {6} \left[ \left( \Sigma_c^{++} - \Sigma_c^0 \right) 
- 2 \left( \Xi^{\prime +}_c -\Xi^{\prime 0}_c \right) \right]$
&& $\frac {\sqrt{3}} 2$ && $2/m_c$ && $1/N_c^2$ && $\epsilon\epsilon^\prime$
&&\omit\cr\space
&  &&  && $\qquad+
\frac 1 6 \left[ \left( \Sigma_c^{*++} - \Sigma_c^{*0} \right) 
- 2 \left( \Xi^{* +}_c -\Xi^{* 0}_c \right) \right]$ 
&& && &&  &&
&&\omit\cr\space\space\space\space\space
& $J_c^i \left(\{T^3,G^{i8} \}-\{T^8,G^{i3} \}\right)$ 
&& $(10 + \overline {10},1)$ && $-{1 \over 2 }
\left[ \Lambda_c^+ \Sigma^+_c - \left( \Xi_c^+ \Xi_c^{\prime +}
- \Xi_c^0 \Xi_c^{\prime 0} \right) \right]$
&& $\frac 3 4$ && $2/m_c$ && $1/N_c^2$ && $\epsilon\epsilon^\prime$
&&\omit\cr\space\space\space\space\space
\space\tablerule\space\space\space
& \omit && \omit &&\hfil $I=2$ \hfil && &&  && && &&\omit\cr
\space\space\space\space
& $\{ T^3, T^3 \}$ && $(27,0)$ && $\frac 1 {10} \left( \Sigma_c^{++}
-2 \Sigma_c^+ + \Sigma_c^0 \right) + \frac 2 5 \left( \Sigma^{*++}_c
-2 \Sigma^{*+}_c + \Sigma^{*0}_c \right)$
&& $2$ && $1$ && $1/N_c$ &&$\epsilon^{\prime\prime}$
&&\omit\cr\space\space\space\space\space
& $J_c^i \{ T^3, G^{i3} \}$ && $(27,1)$ && $-\frac 1 4 \left( \Sigma_c^{++}
-2 \Sigma_c^+ + \Sigma_c^0 \right) + \frac 1 4 \left( \Sigma^{*++}_c
-2 \Sigma^{*+}_c + \Sigma^{*0}_c \right)$
&& $\frac 5 8$ && $2/m_c$ && $1/N_c^2$ &&$\epsilon^{\prime\prime}$
&&\omit\cr\space\space\space\space\space
\space\tablerule
}
\vskip 2pt \hrule
}}
\end{table}

\vfill\break\eject

\begin{table}[htbp]
\caption{Isospin-violating mass splittings of baryons containing a single
bottom quark.}
\smallskip
\label{tab:five}
\centerline{\vbox{ \tabskip=0pt \offinterlineskip
\def\tablerule{\noalign{\hrule}}
\def\space{height 2pt&\omit&&\omit&&\omit&&\omit&&\omit&&\omit&&
\omit&&\omit\cr}
\hrule \vskip 2pt
\halign{
\vrule #&\strut\hfil\ #\hfil&&
\vrule #&\strut\hfil\ #\ \hfil\cr
\tablerule\space
& Operator && $(SU(3), J_b)$ && \hfil  Mass Combination \hfil &&
$\langle {\cal O} \rangle$ && $1/m_b$ && $1/N_c$ && Flavor &&\omit\cr
\tablerule\space\space\space
& \omit && \omit &&\hfil $I=1$ \hfil &&  && && && &&\omit\cr
\space\space\space\space
& $T^3$ && $(8,0)$ && $\left(\Xi_b^0 - \Xi_b^- \right)$  
&& $1$ && $1$  && $1$  && $\epsilon^\prime$ 
&&\omit\cr\space\space\space\space\space
& $J_\ell^i G^{i3}$ && $(8,0)$ && $-\frac 5 8 \left( \Xi^0_b - \Xi^-_b \right) 
+ \frac 1 {24} \left[ 2 \left( \Sigma^+_b - \Sigma^-_b \right)
+ \left( \Xi^{\prime 0}_b - \Xi^{\prime -}_b \right) \right]$  
&& $\frac 5 8$ && $1$ && $1/N_c$ && $\epsilon^\prime$
&&\omit\cr\space
&  &&  && $\qquad + \frac 1 {12} \left[ 2 \left( \Sigma^{*+}_b 
- \Sigma^{*-}_b \right)
+ \left( \Xi^{*0}_b - \Xi^{*-}_b \right) \right]$  
&& &&  &&  && 
&&\omit\cr\space\space\space\space\space
& $J_b^i G^{i3}$ && $(8,1)$ && $\frac 1 2  \left[
\Lambda_b^0\Sigma_b^0 + \frac 1 2 \left( \Xi^0_b \Xi^{\prime 0}_b - \Xi^-_b
\Xi^{\prime -}_b \right) \right]$  
&& $\frac {3\sqrt{3}} {8}$ && $2/m_b$ && $1/N_c$ && $\epsilon^\prime$
&&\omit\cr\space\space\space\space\space
& $\left( J_\ell \cdot J_b \right) T^3$ && $(8,1)$ && $-\frac 1 {6}
\left[ 2 \left(\Sigma_b^+ - \Sigma_b^- \right) + \left(\Xi_b^{\prime 0}
-\Xi_b^{\prime -} \right) \right]$
&& $\frac 5 4$ && $2/m_b$ && $1/N_c^2$ && $\epsilon^\prime$
&&\omit\cr\space
&  &&  && $\qquad -{5 \over {3\sqrt{3}} } \left[
\Lambda_b^0\Sigma_b^0 + \frac 1 2 \left( \Xi^0_b \Xi^{\prime 0}_b - \Xi^-_b
\Xi^{\prime -}_b \right) \right]$ 
&& &&  &&  &&
&&\omit\cr\space
&  &&  && $\qquad\qquad + \frac 1 {6} \left[
2 \left(\Sigma_b^{*+} - \Sigma_b^{*-} \right) 
+ \left(\Xi_b^{* 0} -\Xi_b^{* -} \right) \right] $  
&& &&  &&  && 
&&\omit\cr\space\space\space\space\space
& $\{T^3,T^8\}$ && $(27,0)$ && $
\frac 1 {18} \left[ \left( \Sigma_b^+ - \Sigma_b^- \right) 
- 2 \left( \Xi^{\prime 0}_b -\Xi^{\prime -}_b \right) \right]$
&& $\frac 1 {\sqrt{3}}$ && $1$ && $1/N_c$ && $\epsilon\epsilon^\prime$
&&\omit\cr\space
&  &&  && $\qquad +
\frac 1 9 \left[ \left( \Sigma_b^{*+} - \Sigma_b^{*-} \right) 
- 2 \left( \Xi^{* 0}_b -\Xi^{* -}_b \right) \right]$
&& &&  &&  &&
&&\omit\cr\space\space\space\space\space
& $J_b^i \left(\{T^3,G^{i8} \}+\{T^8,G^{i3} \}\right)$ 
&& $(27,1)$ && $-
\frac 1 {6} \left[ \left( \Sigma_b^+ - \Sigma_b^- \right) 
- 2 \left( \Xi^{\prime 0}_b -\Xi^{\prime -}_b \right) \right]$
&& $\frac {\sqrt{3}} 2$ && $2/m_b$ && $1/N_c^2$ && $\epsilon\epsilon^\prime$
&&\omit\cr\space
& && && $\qquad +
\frac 1 6 \left[ \left( \Sigma_b^{*+} - \Sigma_b^{*-} \right) 
- 2 \left( \Xi^{* 0}_b -\Xi^{* -}_b \right) \right]$
&& &&  &&  &&
&&\omit\cr\space\space\space\space\space
& $J_b^i \left(\{T^3,G^{i8} \}-\{T^8,G^{i3} \}\right)$ 
&& $(10 + \overline {10},1)$ && $-\frac 1 2
\left[ \Lambda_b^0 \Sigma^0_b - \left( \Xi_b^0 \Xi_b^{\prime 0}
- \Xi_b^- \Xi_b^{\prime -} \right) \right]$
&& $\frac 3 4$ && $2/m_b$ && $1/N_c^2$ && $\epsilon\epsilon^\prime$
&&\omit\cr\space\space\space\space\space
\space\tablerule\space\space\space
& \omit && \omit &&\hfil $I=2$ \hfil &&  && && && &&\omit\cr
\space\space\space\space
& $\{ T^3, T^3 \}$ && $(27,0)$ && $\frac 1 {10} \left( \Sigma_b^+
-2 \Sigma_b^0 + \Sigma_b^- \right) + \frac 2 5 \left( \Sigma^{*+}_b
-2 \Sigma^{*0}_b + \Sigma^{*-}_b \right)$
&& $2$ && $1$ && $1/N_c$ &&$\epsilon^{\prime\prime}$
&&\omit\cr\space\space\space\space\space
& $J_b^i \{ T^3, G^{i3} \}$ && $(27,1)$ && $-\frac 1 {4} \left( \Sigma_b^+
-2 \Sigma_b^0 + \Sigma_b^- \right) + \frac 1 4 \left( \Sigma^{*+}_b
-2 \Sigma^{*0}_b + \Sigma^{*-}_b \right)$
&& $\frac 5 8$ && $2/m_b$ && $1/N_c^2$ &&$\epsilon^{\prime\prime}$
&&\omit\cr\space\space\space\space
\space\tablerule
}
\vskip 2pt \hrule
}}
\end{table}

\vfill\break\eject
\begin{table}[htbp]
\caption{Hierarchy of heavy baryon mass splittings in decreasing order of
magnitude for $Q=c$ and $Q=b$.  Superscripts refer to the isospin $I_z$
of the baryon rather than the electromagnetic charge.  The parameters
$(\epsilon^\prime \Lambda_\chi)$ and $(\epsilon^{\prime\prime} \Lambda_\chi)$
are expected to be comparable and of order a few MeV.}
\smallskip
\label{tab:six}
\centerline{\vbox{ \tabskip=0pt \offinterlineskip
\def\tablerule{\noalign{\hrule}}
\def\space{height 2pt&\omit&&\omit&&\omit&&\omit&&\omit\cr}
\hrule \vskip 2pt
\halign{
\vrule #&\strut\hfil\ #\hfil&&
\vrule #&\strut\hfil\ #\ \hfil\cr
\tablerule\space
& \hfil  Mass Combination \hfil &&
Theory && $Q=c$ && $Q=b$ &&\omit\cr
\tablerule\space\space\space
& $\left(\Xi_Q^{+\frac 1 2} - \Xi_Q^{-\frac 1 2} \right)$  
&& $(\epsilon^\prime \Lambda_\chi)$ && $1.0(\epsilon^\prime \Lambda_\chi)$ 
&& $1.0(\epsilon^\prime \Lambda_\chi)$ 
&&\omit\cr\space\space\space\space\space
& $\frac 1 {10} \left( \Sigma_Q^{+1}
-2 \Sigma_Q^0 + \Sigma_Q^{-1} \right) + \frac 2 5 \left( \Sigma^{*+1}_Q
-2 \Sigma^{*0}_Q + \Sigma^{*-1}_Q \right)$
&& $2 {1 \over N_c} (\epsilon^{\prime\prime} \Lambda_\chi)$ && 
$0.67(\epsilon^{\prime\prime} \Lambda_\chi)$ 
&& $0.67 (\epsilon^{\prime\prime} \Lambda_\chi)$
&&\omit\cr\space\space\space\space\space
& $-\frac 5 8 \left( \Xi^{+\frac 1 2}_Q - \Xi^{-\frac 1 2}_Q \right) 
+ \frac 1 {24} \left[ 2 \left( \Sigma^{+1}_Q - \Sigma^{-1}_Q \right)
+ \left( \Xi^{\prime +\frac 1 2}_Q - \Xi^{\prime -\frac 1 2}_Q \right) 
\right]$  
&& $\frac 5 8 {1 \over N_c} (\epsilon^\prime \Lambda_\chi)$ && 
$0.21 (\epsilon^\prime \Lambda_\chi)$ &&
$0.21 (\epsilon^\prime \Lambda_\chi)$ 
&&\omit\cr\space
& $\qquad + \frac 1 {12} \left[ 2 \left( \Sigma^{*+1}_Q 
- \Sigma^{*-1}_Q \right)
+ \left( \Xi^{*+\frac 1 2}_Q - \Xi^{*-\frac 1 2}_Q \right) \right]$  
&& &&  && 
&&\omit\cr\space\space\space\space\space
& $
\frac 1 {18} \left[ \left( \Sigma_Q^{+1} - \Sigma_Q^{-1} \right) 
- 2 \left( \Xi^{\prime +\frac 1 2}_Q -\Xi^{\prime -\frac 1 2}_Q \right) 
\right]$
&& ${1 \over \sqrt{3}} {1 \over N_c} \epsilon (\epsilon^\prime \Lambda_\chi)$ 
&& $0.05 (\epsilon^\prime \Lambda_\chi)$ 
&& $0.05 (\epsilon^\prime \Lambda_\chi)$ 
&&\omit\cr\space
& $\qquad +
\frac 1 9 \left[ \left( \Sigma_Q^{*+1} - \Sigma_Q^{*-1} \right) 
- 2 \left( \Xi^{* +\frac 1 2}_Q -\Xi^{* -\frac 1 2}_Q \right) \right]$
&& &&  &&
&&\omit\cr\space\space\space\space\space
& $\frac 1 2  \left[
\Lambda_Q^0\Sigma_Q^0 + \frac 1 2 \left( \Xi^{+\frac 1 2}_Q 
\Xi^{\prime +\frac 1 2}_Q - \Xi^{-\frac 1 2}_Q
\Xi^{\prime -\frac 1 2}_Q \right) \right]$  
&& ${{3\sqrt{3}} \over 8} {1\over N_c}{{2\Lambda} \over m_Q} (\epsilon^\prime \Lambda_\chi)$
&& $0.09(\epsilon^\prime \Lambda_\chi)$
&& $0.03(\epsilon^\prime \Lambda_\chi)$
&&\omit\cr\space\space\space\space\space
& $-\frac 1 {6}
\left[ 2 \left(\Sigma_Q^{+1} - \Sigma_Q^{-1} \right) + 
\left(\Xi_Q^{\prime +\frac 1 2}
-\Xi_Q^{\prime -\frac 1 2} \right) \right]$
&& $\frac 5 4 {1 \over N_c^2}{{2\Lambda} \over m_Q} (\epsilon^\prime \Lambda_\chi)$
&& $0.06(\epsilon^\prime \Lambda_\chi)$ 
&& $0.02(\epsilon^\prime \Lambda_\chi)$ 
&&\omit\cr\space
& $\qquad -{5 \over {3\sqrt{3}} } \left[
\Lambda_Q^0\Sigma_Q^0 
+ \frac 1 2 \left( \Xi^{+\frac 1 2}_Q \Xi^{\prime +\frac 1 2}_Q 
- \Xi^{-\frac 1 2}_Q \Xi^{\prime -\frac 1 2}_Q \right) \right]$ 
&& &&  &&
&&\omit\cr\space
& $\qquad\qquad + \frac 1 {6} \left[
2 \left(\Sigma_Q^{*+1} - \Sigma_Q^{*-1} \right) 
+ \left(\Xi_Q^{* +\frac 1 2} -\Xi_Q^{* -\frac 1 2} \right) \right] $  
&& &&  &&  
&&\omit\cr\space\space\space\space\space
& $-\frac 1 {4} \left( \Sigma_Q^{+1}
-2 \Sigma_Q^0 + \Sigma_Q^{-1} \right) + \frac 1 4 \left( \Sigma^{*+1}_Q
-2 \Sigma^{*0}_Q + \Sigma^{*-1}_Q \right)$
&& $\frac 5 8 {1 \over N_c^2} {{2\Lambda} \over m_Q} (\epsilon^{\prime\prime} \Lambda_\chi)$
&& $0.03(\epsilon^{\prime\prime} \Lambda_\chi)$
&& $0.009(\epsilon^{\prime\prime} \Lambda_\chi)$
&&\omit\cr\space\space\space\space
& $-
\frac 1 {6} \left[ \left( \Sigma_Q^{+1} - \Sigma_Q^{-1} \right) 
- 2 \left( \Xi^{\prime +\frac 1 2}_Q 
-\Xi^{\prime -\frac 1 2}_Q \right) \right]$
&& $\frac {\sqrt{3}} 2 {1 \over N_c^2} 
{{2\Lambda} \over m_Q} \epsilon(\epsilon^\prime \Lambda_\chi)$ 
&& $0.01(\epsilon^\prime \Lambda_\chi)$ 
&& $0.003(\epsilon^\prime \Lambda_\chi)$ 
&&\omit\cr\space
& $\qquad +
\frac 1 6 \left[ \left( \Sigma_Q^{*+1} - \Sigma_Q^{*-1} \right) 
- 2 \left( \Xi^{* +\frac 1 2}_Q -\Xi^{* -\frac 1 2}_Q \right) \right]$
&& &&  &&
&&\omit\cr\space\space\space\space\space
& $-\frac 1 2
\left[ \Lambda_Q^0 \Sigma^0_Q - \left( \Xi_Q^{+\frac 1 2} 
\Xi_Q^{\prime +\frac 1 2}
- \Xi_Q^{-\frac 1 2} \Xi_Q^{\prime -\frac 1 2} \right) \right]$
&& $\frac 3 4 {1 \over N_c^2} 
{{2 \Lambda} \over m_Q} \epsilon(\epsilon^\prime \Lambda_\chi)$
&& $0.009(\epsilon^\prime \Lambda_\chi)$
&& $0.0026(\epsilon^\prime \Lambda_\chi)$
&&\omit\cr\space\space\space\space\space
\space\tablerule
}
\vskip 2pt \hrule
}}
\end{table}

\vfill\break\eject

\end{document}